\documentclass[pdflatex,sn-mathphys-num]{sn-jnl}
\usepackage{fontspec}
\usepackage{comment}
\usepackage{graphicx}
\usepackage{multirow}
\usepackage{amsmath,amssymb,amsfonts}
\usepackage{amsthm}
\usepackage{mathrsfs}
\usepackage[title]{appendix}
\usepackage{xcolor}
\usepackage{textcomp}
\usepackage{manyfoot}
\usepackage{booktabs}
\usepackage{algorithm}
\usepackage{algorithmicx}
\usepackage{algpseudocode}
\usepackage{listings}
\usepackage{subcaption}
\usepackage{url}
\usepackage{hyperref}

\theoremstyle{thmstyleone}

\theoremstyle{thmstyletwo}

\theoremstyle{thmstylethree}

\begin{document}

\title[Individual Fairness in Community Detection: Quantitative Measure and Comparative Evaluation]{Individual Fairness in Community Detection: Quantitative Measure and Comparative Evaluation}

\author[]{\fnm{Fabrizio} \sur{Corriera}}\email{f.corriera@liacs.leidenuniv.nl}

\author[]{\fnm{Frank W.} \sur{Takes}}\email{f.w.takes@liacs.leidenuniv.nl}

\author[]{\fnm{Akrati} \sur{Saxena}}\email{a.saxena@liacs.leidenuniv.nl}

\affil[]{\orgdiv{LIACS}, \orgname{Leiden University}, \orgaddress{\country{the Netherlands}}}

\abstract{
Community detection is a fundamental task in complex network analysis. Fairness-aware community detection seeks to prevent biased node partitions, typically framed in terms of individual fairness, which requires similar nodes to be treated similarly, and group fairness, which aims to avoid disadvantaging specific groups of nodes. 
While existing literature on fair community detection has primarily focused on group fairness, we introduce a novel measure to quantify individual fairness in community detection methods. The proposed measure captures unfairness as the vectorial distance between a node’s true and predicted community representations, computed using the community co-occurrence matrix.
We provide a comprehensive empirical investigation of a broad set of community detection algorithms from the literature on both synthetic networks, with varying levels of community explicitness, and real-world networks. We particularly investigate the fairness-performance trade-off using standard quality metrics and compare individual fairness outcomes with existing group fairness measures. The results show that individual unfairness can occur even when group fairness or clustering accuracy is high, underscoring that individual and group fairness are not interchangeable. Moreover, fairness depends critically on the detectability of community structure. However, we find that Significance and Surprise for denser graphs, and Combo, Leiden, and SBMDL for sparser graphs result in a better trade-off between individual fairness and community quality. Overall, our findings, together with the fact that community detection is an important step in many network analysis downstream tasks, highlight the necessity of developing fairness-aware community detection methods.
}


\maketitle

\section{Introduction}\label{sec:introduction}

Community detection (CD) is a fundamental task in network science that seeks to uncover the underlying modular structure of networks by grouping together nodes that are more densely connected to one another than to the rest of the network~\cite{Fortunato2009}. Identifying such communities is crucial for analysing social networks, where communities often correspond to groups with shared interests, social roles, or demographic characteristics~\cite{chakraborty2017metrics}. Accurately detected communities provide a deeper understanding of underlying social structures~\cite{menyhert2025connectivity}. CD is furthermore used as an integral part of other downstream tasks in network analysis, such as link prediction, influence maximization, and recommendation systems~\cite{saxena2024fairsnaalgorithmicfairnesssocial}.

Despite its widespread use, CD methods can exacerbate or reproduce biases in social networks \cite{saxena2024fairsnaalgorithmicfairnesssocial}. Social networks encode pre-existing structural inequalities arising due to many factors, such as homophily, minority under-representation, and structural segregation~\cite{saxena2024fairsnaalgorithmicfairnesssocial, saxena2025homophily}. If these inequalities are not considered in designing a CD method, the method might not accurately identify communities. For example, when marginalised groups have fewer connections across communities, traditional fairness-agnostic CD algorithms may disproportionately assign them to fragmented or peripheral clusters \cite{mehrabi2019debiasing}. This not only affects how their roles are represented in the network but also influences subsequent downstream network analysis algorithms (link prediction, centrality ranking, influence maximisation, etc.) that rely on these community assignments. Biased detected communities may amplify disparities in access to information, or reinforce systemic biases in network analysis-based applications~\cite{saxena2024fairsnaalgorithmicfairnesssocial}.

Although many CD algorithms have been proposed in the literature~\cite{chakraborty2017metrics, jin2021surveycommunitydetectionapproaches}, most overlook the structural inequalities inherent in real-world networks~\cite{saxena2024fairsnaalgorithmicfairnesssocial, devink2024groupfairnessmetricscommunity}, as most CD methods focus on optimizing clustering quality. In literature, various metrics exist to assess the quality of detected communities~\cite{chakraborty2017metrics}, including modularity~\cite{newman2004finding}, Normalized Mutual Information (NMI)~\cite{lancichinetti2009detecting}, and Adjusted Rand Index (ARI)~\cite{ari}. However, metrics to evaluate the fairness of identified communities remain comparatively underexplored. Prior work on fairness in CD~\cite{devink2024groupfairnessmetricscommunity, panayiotou2024fair, Gkartzios_group_modularity, de2025measuring} has primarily focused on group fairness, ensuring that groups defined by sensitive attributes are fairly represented.
However, the existing metrics fail to capture individual fairness, i.e., whether similar individuals are treated similarly~\cite{carey2022fairness}. This raises a critical question: how can we evaluate and guarantee fairness at the individual level in Community detection?

In this paper, we address the thusfar unexplored problem of individual fairness in CD, proposing a novel measure to capture it. Our proposed measure, Individual Bias ($IB$), quantifies the unfairness of a given CD method at the node level, based on nodes' community (or ``context") changes. $IB$ captures node-level unfairness by quantifying how much a node’s predicted context deviates from its true one. This is achieved using the Community Co-occurrence (CC) matrix, which records how often each node appears in the same community as every other node, giving us a representation of what we refer to as local context. Individual Bias of a node is computed as a vectorial distance between the ground-truth and predicted co-occurrence vectors. This comparison reflects how well a node’s context is preserved, where lower $IB$ values indicate fair placement, while higher values reveal substantial distortion in the node’s community affiliation. Building on our node-level measure, we propose a graph-level extension, $IB_G$, computed as the standard deviation of the node-level $IB$ scores. It provides a comprehensive view of individual unfairness, indicating whether all nodes are treated equally fairly or not.

We evaluate the proposed measure on both synthetic networks generated using the ABCD benchmark~\cite{Kaminski2021} and multiple real-world networks, applying 30 widely used CD algorithms. For each algorithm, we compare fairness with standard metrics to compute the quality of identified communities. Our results show that no single algorithm or a specific class of algorithms perform uniformly well across all datasets. We observe that, for denser graphs, the CD algorithms Significance and Surprise offer a good trade-off between individual fairness and the quality of the predicted communities. On the contrary, for sparser graphs, the best trade-off is shown by the algorithms Combo, Leiden, and SBMDL. Comparing individual fairness with group fairness, our experiments highlight that many CD methods have high individual bias even when group-level fairness or overall CD accuracy is high, underscoring the necessity of considering fairness at the individual level to remove biases towards already marginalised individuals.

The remainder of the paper is structured as follows. Section~\ref{sec:related_work} reviews related work on fair CD and individual fairness in social network analysis. Section~\ref{sec:preliminaries} introduces definitions and data structures for our approach. Section~\ref{sec:proposed_fairness_measures} defines the proposed Individual Bias measure. Section~\ref{sec:data_experimental_setup} presents the experimental setup, including datasets, CD methods, and evaluation metrics. Section~\ref{sec:results} reports results and analyses fairness trends across algorithms. Section~\ref{sec:conclusion} concludes our work and gives future research directions.

\section{Related work}\label{sec:related_work}

In this section, we first present recent works on fairness in CD. Then, we discuss individual fairness in network science downstream tasks, particularly, individual fairness in influence maximization. %we then present a study on classification fairness, offering valuable insights into individual fairness and its relationship with group fairness

Although CD has received considerable attention in the literature, the notion of fairness remains largely overlooked, as can be deduced by the absence of well-established metrics to quantify it~\cite{saxena2024fairsnaalgorithmicfairnesssocial}. Most existing CD methods focus on structural accuracy, often overlooking how biases in network structure or algorithmic design may disadvantage specific groups or individuals~\cite{mehrabi2019debiasing}. In this work, it is observed that well-known CD methods often fail to assign lowly connected nodes to proper communities or assign them to numerous small, less meaningful ones. The authors proposed a method, called CLAN, that addresses this by moving such nodes to larger communities using external node attributes not considered in the structural detection process. While CLAN improves assignments for marginalized nodes, it does not explicitly define fairness constraints.

In recent work, de Vink et al.~\cite{devink2024groupfairnessmetricscommunity} proposed a novel class of group fairness metrics that evaluate whether CD methods exhibit systematic bias towards specific types of communities, such as minority groups, or communities with specific structural properties based on their size, density, or conductance. Their framework maps detected communities to the ground-truth and computes community-wise performance using three proposed measures, including (i) Fraction of Correctly Classified Nodes (FCCN), (ii) F1 score, and (iii) Fraction of Correctly Classified Edges (FCCE). Finally, the group fairness ($\Phi$) is computed by fitting these community-wise fairness scores to the normalised community properties (such as size, density, or conductance). The authors compared several CD methods on real-world and synthetic benchmark networks, and recommended a set of CD methods that can be used to obtain fair communities, including Walktrap, Significance, and Infomap.

Gkartzios et al.~\cite{Gkartzios_group_modularity} introduced the concept of fairness using the group modularity, focusing on the connectivity between nodes belonging to different sensitive groups, compared to connections within the same group. Sensitive groups are defined based on sensitive attributes, such as gender, ethnicity, or age. The key question addressed is whether nodes belonging to different sensitive groups are equally well-connected within each community. This approach shifts the focus from balancing the proportion of groups within a community to examining the internal connections (edges) of those groups. The authors proposed two variations of the traditional modularity metric~\cite{newman2004finding}, which quantify the strength of a network's community division by comparing the density of internal edges to the expected density in a random graph. Their new modularity approach specifically looks at the density of edges of nodes belonging to a particular group. These fairness-aware modularity functions were evaluated by using them as the optimization function for the Louvain algorithm~\cite{Blondel_2008}, on both synthetic and real-world networks. The results indicate that higher homophily correlates with increased unfairness and reduced diversity. Moreover, fairness-aware algorithms improve fairness objectives, but this often comes at the cost of lower overall modularity and a larger number of detected communities.

In contrast to group fairness, individual fairness in CD remains largely unexplored. Lessons can be drawn, however, from fairness research in machine learning and social networks more broadly. The foundational principle of fairness through awareness proposed by Dwork et al.~\cite{dwork2011fairnessawareness} asserts that similar individuals should receive similar outcomes. This concept was motivated by concerns that simply achieving fairness between groups might still lead to unfair outcomes at the individual level. Generally speaking, according to the authors, individual fairness requires defining a distance metric based on their task-relevant attributes, ensuring that outcomes respect similarity relations~\cite{dwork2011fairnessawareness}. This means that, given a specific task, the attributes used to define similarity must be chosen coherently with the said task. This way, the desired outcome can be enforced by defining a distance metric that allows unbiased measurement of similarity between individuals. While conceptually appealing, this approach faces practical challenges: defining ``similarity” is highly context-dependent, and constructing a universally accepted distance function is often infeasible. Moreover, some scholars argue that the philosophical underpinnings of individual and group fairness are not inherently conflicting, suggesting that any perceived conflict is an ``artifact of the failure to fully articulate assumptions"~\cite{carey2022fairness}.

In recent years, several studies have examined group fairness in network analysis tasks, such as link prediction~\cite{saxena2022nodesim, saxena2022hm}, influence maximization~\cite{rahmattalabi2021fair, ma2024fair, saxena2025dq4fairim}, influence blocking~\cite{saxena2023fairness}, centrality ranking \cite{tsioutsiouliklis2021fairness, kumar2026fairness}, and opinion dynamics \cite{stkepien2026fairness}. However, individual fairness is not well explored.
Fish et al.~\cite{Fish_2019} investigated individual fairness for influence maximization in terms of information access. They formalized the notion of an information access gap and sought to minimize disparities by maximizing the minimum probability of information reaching each individual. They further proposed four fairness-aware algorithms: Greedy, Myopic, Naive Myopic, and Gonzalez, for influence maximization. Results showed that the Myopic and Naive Myopic perform best in improving outreach and reducing inequities. Their findings highlight that ensuring equitable access to information is crucial for social interventions, awareness campaigns, and public health communication. 

In summary, while fairness in CD has been addressed at the group level, the question of whether individual nodes are treated fairly in the CD remains open. Therefore, in this paper, we propose an individual fairness measure for CD that extends fairness analysis beyond groups to the level of individual nodes.

\section{Preliminaries}\label{sec:preliminaries}
In Section~\ref{subsec:notation_definitions}, we introduce the necessary definitions and data structures used to define our node-level and graph-level individual fairness measures, as explained in Sections~\ref{subsec:IB} and~\ref{subsec:GIB}, respectively. We first discuss the notations (summarized in Table~\ref{tab:notation}), followed by the metrics, explained in~\ref{subsec:vectorial_distance}.

\begin{table}[!t]
    \centering
    \begin{tabular}{l c}
    \toprule
        \textbf{Symbol} & \textbf{Explanation}\\
    \midrule
        $G$ & Graph\\

        $V$ & Set of vertices\\

        $E$ & Set of edges\\

        $\Gamma$ & Community Co-occurrence ($CC$) Matrix for ground-truth communities\\

        $\Gamma'$ & Community Co-occurrence ($CC$) Matrix for predicted communities\\

        $\Gamma_i$ & $i$-th row of matrix $\Gamma$\\

        $\Gamma_i'$ & $i$-th row of matrix $\Gamma'$\\

        $\Gamma_{i,j}$ & Element of matrix $\Gamma$ in row $i$ and column $j$\\

        $\Gamma_{i,j}'$ & Element of matrix $\Gamma'$ in row $i$ and column $j$\\

        $c_i$ & Node $i$'s community\\

        $IB_i$ & Individual bias for node $i$\\

        $IB_G$ & Individual bias in graph G\\
    \bottomrule
    \end{tabular}
    \caption{Notations used throughout the paper.}
    \label{tab:notation}
\end{table}

\subsection{Notation and Definitions}\label{subsec:notation_definitions}

We consider an unweighted undirected graph $G=(V,E)$, where $V$ is the set of its vertices and $E$ is the set of its edges. Each node $i \in V$ belongs to exactly one community in $C$, where $C$ denotes the set of all communities in the graph, and the community membership of node $i$ is denoted by $c_i$.

To capture community membership relationships, we define the CC matrix $\Gamma \in \{0,1\}^{|V|\times|V|}$, where each entry $\Gamma_{i,j}$ indicates whether two nodes $i$ and $j$ belong to the same community or not:

$$
\Gamma_{i,j}=
\begin{cases}
    1, & \text{if }c_i =c_j \\
    0, & \text{if }c_i \neq c_j
\end{cases}
, \forall i,j \in V
$$

Thus, row (or column) $i$ of $\Gamma$ represents the community membership vector of node $i$, indicating which nodes belong to the same community as $i$.

We first construct the CC matrix for the ground-truth communities, which is represented by $\Gamma$. Once a CD method is applied, we obtain a set of predicted communities $C'$, and, analogously, construct a $CC$ matrix $\Gamma'$ based on the assignments of predicted communities.

\subsection{Vectorial Distance Measurement}\label{subsec:vectorial_distance}

To compute our measure, we use the concept of vectorial distance to measure the context variation between the ground truth and the identified communities of the nodes. The methodology of our measure is fully explained in Section~\ref{subsec:IB}. Here, we discuss various properties that should be considered while choosing a distance function.
\begin{itemize}
    \item \textbf{Scale sensitivity:} The function should not be affected by the magnitude of the vectors. The desired behaviour for the measure is that results from networks of different sizes should be comparable. Therefore, scale sensitivity is not a desired quality for the function, and we will look for a scale-invariant vectorial distance function.
    \item \textbf{Direction sensitivity:} The function should capture the differences in the vectors' directions. In contrast to scale sensitivity, this is a crucial quality for comparing the context change of communities. The reason is that the main difference between the vectors will be reflected in their directions rather than their magnitudes.
    \item \textbf{Applicability to real values:} Although the vectors to which we will apply the function are binary vectors, it is desirable that the function generalizes to real-valued vectors to allow for potential extensions of our framework.
    \item \textbf{Differentiability:} Differentiability shows that the distance function is smooth and differentiable. As above, this property is preferred to ensure that the function can be extended and used for other purposes, e.g., as an optimization function.
    \item \textbf{Defined measurement scale:} A bounded and interpretable output range is advantageous, so that it would be easy to compare the outputs of the function without any normalization.
\end{itemize}

Based on these criteria, we examine several widely used vectorial distance measures, including Cosine distance, L1 distance, L2 distance, Dot product distance, Hamming distance, Jaccard distance, and Earth Mover’s Distance (EMD). Their properties, with respect to the above criteria, are summarized in Table~\ref{tab:vectorial_distances_characteristics}.

\begin{table}[t]
    \centering
    \begin{tabular}{l c c c c c}
        \toprule
         & \textbf{Scale invariant} & \textbf{Direction sensitive} & \textbf{Real values} & \textbf{Differentiable} & \textbf{Scale} \\
        \midrule
        Cosine & \checkmark & \checkmark & \checkmark & \checkmark & $[0,2]$ \\

        L1 & $\times$ & $\times$ & \checkmark & $\times$ & $[0,\infty)$ \\

        L2 & $\times$ & $\times$ & \checkmark & \checkmark & $[0,\infty)$ \\

        Dot & $\times$ & \checkmark & \checkmark & \checkmark & $(-\infty,\infty)$ \\

        Hamming & \checkmark & $\times$ & $\times$ & $\times$ & $[0,n]$\\

        Jaccard & \checkmark & $\times$ & $\times$ & $\times$ & $[0,1]$ \\

        EMD & $\times$ & \checkmark & \checkmark & $\times$ & $[0,\infty)$ \\
        \bottomrule
    \end{tabular}
    \caption{Summary of the vectorial distances' characteristics.}
    \label{tab:vectorial_distances_characteristics}
\end{table}

There is no single ``perfect” distance function, and the choice strictly depends on the specific objectives of the analysis. Therefore, the properties summarized in Table~\ref{tab:vectorial_distances_characteristics} should not be interpreted as ``absolute" qualities, since different applications may prioritize different criteria. For the purposes of this work, we select the Cosine distance, as it satisfies all the requirements we outlined.

\subsubsection{Cosine distance}\label{subsubsec:cosine}
Cosine distance quantifies the dissimilarity between two vectors by measuring the angle between them. Given two vectors $v_i$ and $v_j$, their cosine distance is computed as

$$d_{cosine}(v_i,v_j)=1-\frac{v_i \cdot v_j}{\|v_i\| \|v_j\|}$$

Here, we define cosine distance as one minus the cosine similarity. Cosine similarity is normalised by the product of the vector norms, ensuring that the measure is invariant to vector magnitude.

\section{Individual Bias Measures}\label{sec:proposed_fairness_measures}

In Section~\ref{subsec:IB}, we explain the proposed Individual Bias measure, followed by its extension at the graph level in Section~\ref{subsec:GIB}.

\subsection{Node Individual Bias}\label{subsec:IB}

Consider a network $G = (V, E)$ with ground-truth communities $C$ and predicted communities $C'$, obtained from a CD algorithm. Using these assignments, we construct the ground-truth $CC$ matrix $\Gamma$ and the predicted $CC$ matrix $\Gamma'$, respectively.

\begin{figure}[!t]
    \centering
    \includegraphics[trim={0cm 9cm 0cm 3.3cm}, clip, width=\linewidth]{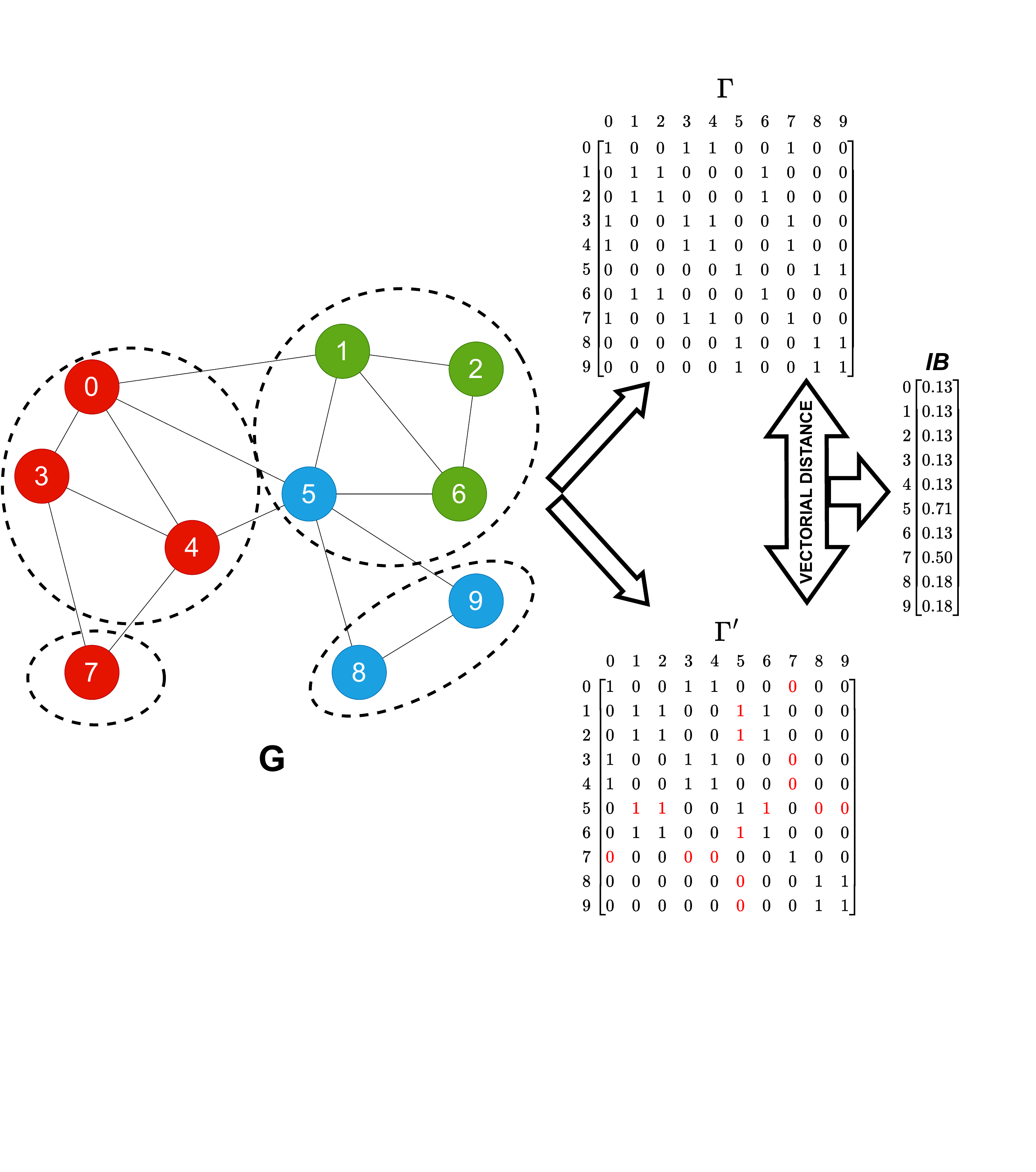}
    \caption{The $IB$ calculation process. On the left is shown a graph with ground-truth communities (coded by colour) and communities identified by a CD method (represented by the dotted ellipses). We first compute the $CC$ matrix using the ground-truth communities and the predicted communities (respectively on top and bottom of the image). We then obtain $IB$ (on the right), applying the vectorial distance between the rows of the two matrices. The values shown are obtained using the cosine distance.}
    \label{fig:graph_2_cc_example}
\end{figure}

As explained in Section~\ref{subsec:notation_definitions}, each row of $\Gamma$ encodes the local context of a node, describing with which other nodes it shares a community. Therefore, we could quantify the dissimilarity between $\Gamma_i$ and $\Gamma'_i$, to measure how accurately a node’s community membership has been preserved. As we anticipated in~\ref{subsec:vectorial_distance}, we use a vectorial distance function to compute $IB$, and it is used to quantify the difference between ground-truth community affiliation vectors and predicted community affiliation vectors. If a CD method identifies communities accurately, the local context of each node remains unchanged. However, if the method systematically misclassifies certain types of nodes, the discrepancy between their ground-truth and predicted co-occurrence patterns will be larger. Moreover, these errors propagate, since misplacing one node also alters the fairness of its neighbours’ co-occurrence patterns. Fig.~\ref{fig:graph_2_cc_example} summarises the process just described.

We quantify this context change at the node level through the Individual Bias ($IB$). For a node $i \in V$, $IB_i$ is defined as the cosine distance between its ground-truth and predicted co-occurrence vectors:

$$IB_i= d_{cosine}(\Gamma_i,\Gamma'_i)=1-\frac{\Gamma_i \cdot \Gamma'_i}{\|\Gamma_i\| \|\Gamma'_i\|}$$

Cosine distance is scale-invariant, sensitive to direction, works with real values vectors, it is differentiable and its scale is $[0,2]$, but since we are only working with positive values, the quantity $\frac{\Gamma_i \cdot \Gamma'_i}{\|\Gamma_i\| \|\Gamma'_i\|}$ will never be negative, so, the actual scale for cosine distance, applied to our problem, is $[0,1]$.

$IB_i = 0$ indicates perfect preservation of node $i$’s context (fair treatment). Higher values of $IB_i$ reflect greater deviation between predicted and true co-occurrence patterns, highlighting individual unfairness.

\subsection{Range of Individual Bias}\label{subsec:range_ib}

In our hypothesis, since $\Gamma_i$ and $\Gamma'_i$ are binary vectors, the dot product in the numerator satisfies: $\Gamma_i \cdot \Gamma'_i\le\sum_{j=1}^{|V|}\Gamma_{i,j}$. This is because each entry in each vector can only be equal to $1$ or $0$, and our best case scenario is when $\Gamma_i=\Gamma'_i$. Intuitively, expanding the predicted community (i.e., adding nodes incorrectly) increases the denominator without increasing the numerator, while shrinking it (i.e., missing true members) decreases both, typically reducing the similarity.

As stated in~\ref{subsubsec:cosine}, the range of $IB$ is $[0,1]$, with $0$ meaning a perfectly fair community prediction for a certain node, and $1$ being a perfectly unfair community prediction with respect to a given node $i$. However, it is important to note that, under our assumptions, not both the extreme values of the function’s range are attainable. Specifically, it is impossible for $\Gamma'_i = \neg \Gamma_i$, since each node necessarily belongs to its own community and therefore $\Gamma_{i,i}=1, \forall i$.

The same property holds for any predicted co-occurrence matrix $\Gamma'$ constructed from a CD algorithm. This means that:

\begin{equation}
    \label{eq:CC_diagonal_equality}
    \Gamma_{i,i}=\Gamma'_{i,i}=1,\forall i \in V
\end{equation}

Therefore, assuming $\Gamma'i = \neg \Gamma_i$ would require $\Gamma{i,i} = 0$, which contradicts the community-belonging property in Equation~\ref{eq:CC_diagonal_equality}. In other words, such a situation would imply that a node does not belong to its own community, which is impossible.

Consequently, the effective range of our function is the semi-open interval $[0,1)$, even though, given a large enough graph, absolute unfair measures might get very close to $1$.

\subsection{Graph Individual Bias}\label{subsec:GIB}

While $IB$ measures how fairly a single node is treated by a CD algorithm, we extend this measure to compute graph-level bias, assessing the algorithm's overall individual fairness across the entire network.

For an algorithm to be considered individually fair at the graph level, all nodes should experience a similar degree of fairness. In other words, the $IB$ values should be tightly concentrated around a common value. To capture this, we introduce the Graph Individual Bias ($IB_G$), computed as the standard deviation of the node-level $IB$ scores across all nodes. Thus,

\begin{equation}
    \label{eq:std_IB_u}
    IB_G=\sigma(\{IB_i, \forall i \in V\})
\end{equation}

Since $IB_i \in [0,1)$, the standard deviation $IB_G$ ranges between $0$ and $0.5$.

A value of $IB_G \approx 0$ indicates that all nodes are treated similarly, meaning the algorithm behaves uniformly across the network, treating all nodes in an equally fair or unfair manner.
Higher values of $IB_G$ indicate a greater disparity across nodes, revealing uneven treatment and, therefore, higher unfairness at the graph level. In this way, $IB_G$ complements node-level $IB$ by providing a global perspective on how consistently a CD algorithm treats all nodes in the network.

\subsection{Behavioural Analysis of Individual Bias}\label{subsec:behav_analysis_IB}
In this section, we study the behaviour of the $IB$ measure under different scenarios. To do so, we generate synthetic networks with ground-truth communities and analyse how a node’s context changes under three different scenarios:

\begin{itemize}
    \item Context expansion
    \item Context shrinkage
    \item Context change
\end{itemize}

We generate networks of three different sizes: 100, 1,000, and 10,000 nodes. Each network has two communities: a minority community comprising $20\%$ of the nodes and a majority community comprising $80\%$, referred to as $c_m$ and $c_M$, respectively. All reported plots show the mean values over 100 random runs.

\begin{figure}[t]
    \centering
    \begin{subfigure}[b]{.32\linewidth}
        \begin{center}
            100 nodes
        \end{center}
        \includegraphics[width=\linewidth]{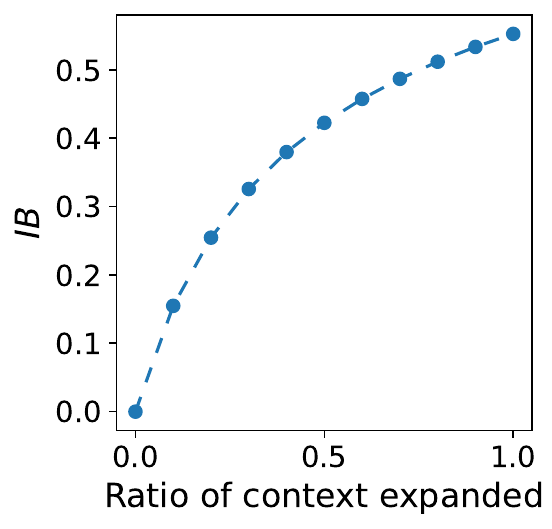}
        \caption{}
        \label{fig:expanded_02_100}
    \end{subfigure}
    \begin{subfigure}[b]{.32\linewidth}
        \begin{center}
            1,000 nodes
        \end{center}
        \includegraphics[width=\linewidth]{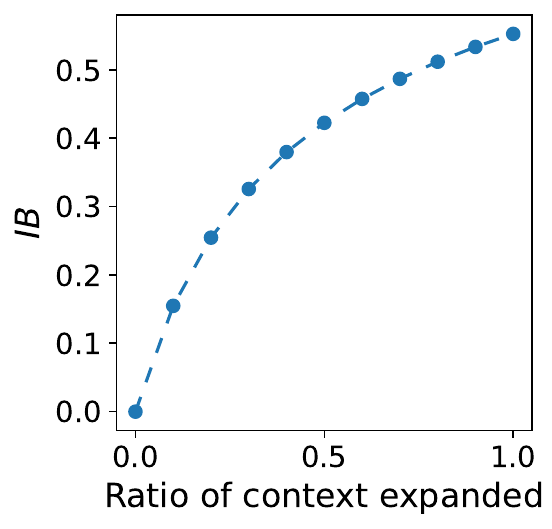}
        \caption{}
        \label{fig:expanded_02_1000}
    \end{subfigure}
    \begin{subfigure}[b]{.32\linewidth}
        \begin{center}
             10,000 nodes
        \end{center}
        \includegraphics[width=\linewidth]{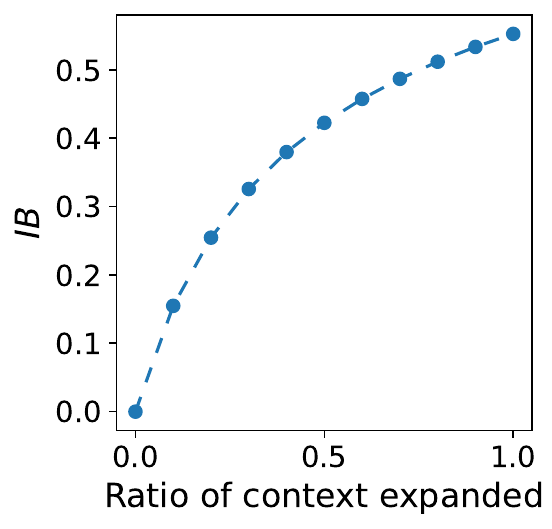}
        \caption{}
        \label{fig:expanded_02_10000}
    \end{subfigure}
    \begin{subfigure}[b]{.32\linewidth}
        \includegraphics[width=\linewidth]{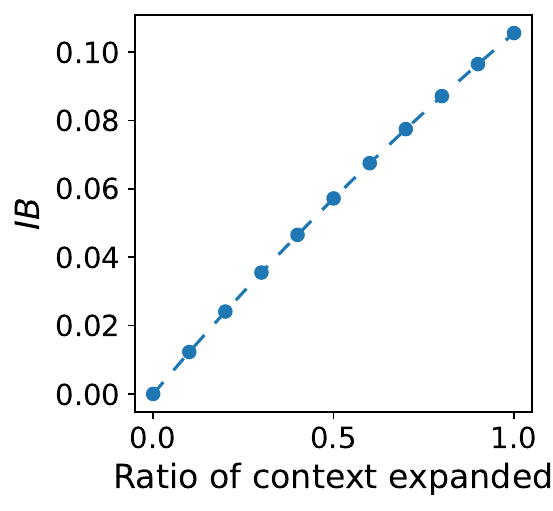}
        \caption{}
        \label{fig:expanded_08_100}
    \end{subfigure}
    \begin{subfigure}[b]{.32\linewidth}
        \includegraphics[width=\linewidth]{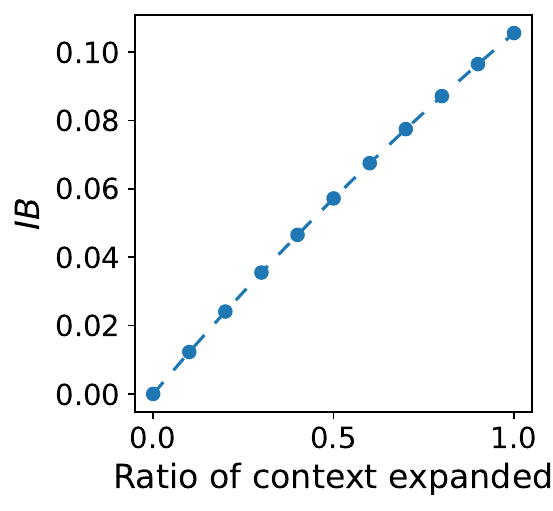}
        \caption{}
        \label{fig:expanded_08_1000}
    \end{subfigure}
    \begin{subfigure}[b]{.32\linewidth}
        \includegraphics[width=\linewidth]{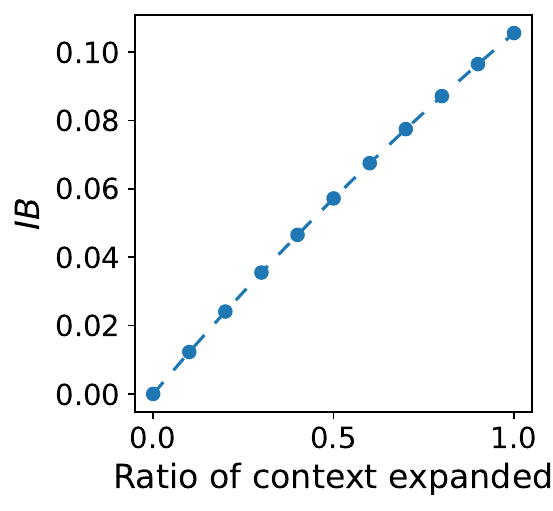}
        \caption{}
        \label{fig:expanded_08_10000}
    \end{subfigure}
    \caption{$IB$ behaviour of the \textbf{expanding} context of a node belonging to the minority community ((a), (b) and (c)) and majority community ((d), (e) and (f)). (a) and (d) correspond to graphs with 100 nodes, (b) and (e) correspond to graphs with 1,000 nodes, and (d) and (f) correspond to graphs with 10,000 nodes.}
    \label{fig:expanded}
\end{figure}

Figure~\ref{fig:expanded} shows how the individual bias $IB$ of a node changes when its community is expanded, meaning the predicted community is larger than the ground-truth one. In Figures~\ref{fig:expanded_02_100},~\ref{fig:expanded_02_1000}, and~\ref{fig:expanded_02_10000} node belongs to $c_m$, while, in plots~\ref{fig:expanded_08_100},~\ref{fig:expanded_08_1000}, and~\ref{fig:expanded_08_10000}, it belongs to $c_M$. 
All figures exhibit a concave-down increasing trend, but the curve for $c_M$ is almost linear. Notably, the magnitude of $IB$ for the majority community ($c_M$) is much smaller (approximately $0.10$ compared to $0.5$ for the minority ($c_m$)). This difference arises from the relative community sizes: adding nodes to $c_M$ produces a proportionally smaller change in its context than adding nodes to $c_m$, resulting in the steeper curve observed for the minority community.

\begin{figure}[t]
    \centering
    \begin{subfigure}[b]{.32\linewidth}
        \begin{center}
            100 nodes
        \end{center}
        \includegraphics[width=\linewidth]{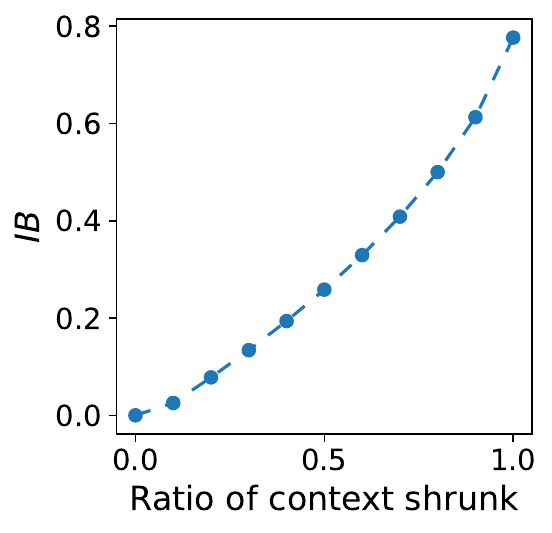}
        \caption{}
        \label{fig:shrunk_02_100}
    \end{subfigure}
    \begin{subfigure}[b]{.32\linewidth}
        \begin{center}
            1,000 nodes
        \end{center}
        \includegraphics[width=\linewidth]{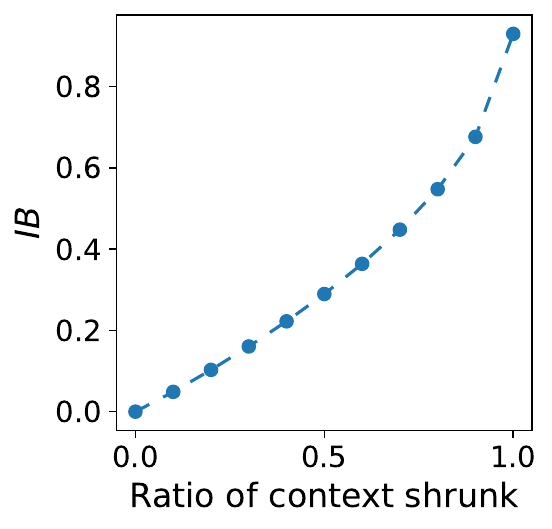}
        \caption{}
        \label{fig:shrunk_02_1000}
    \end{subfigure}
    \begin{subfigure}[b]{.32\linewidth}
        \begin{center}
             10,000 nodes
        \end{center}
        \includegraphics[width=\linewidth]{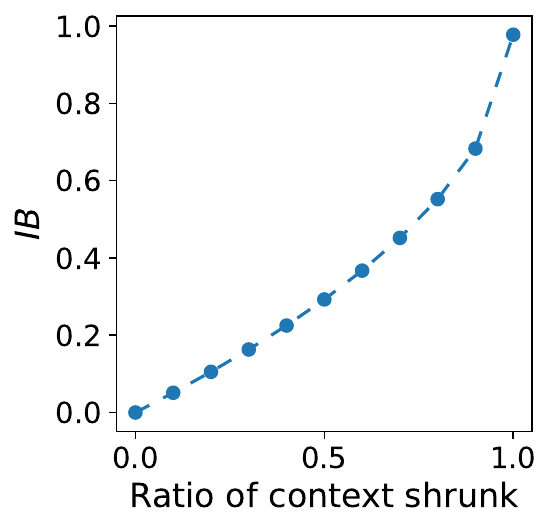}
        \caption{}
        \label{fig:shrunk_02_10000}
    \end{subfigure}
    \begin{subfigure}[b]{.32\linewidth}
        \includegraphics[width=\linewidth]{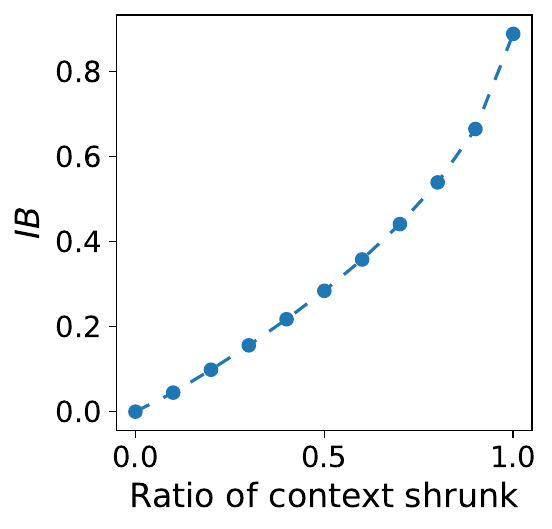}
        \caption{}
        \label{fig:shrunk_08_100}
    \end{subfigure}
    \begin{subfigure}[b]{.32\linewidth}
        \includegraphics[width=\linewidth]{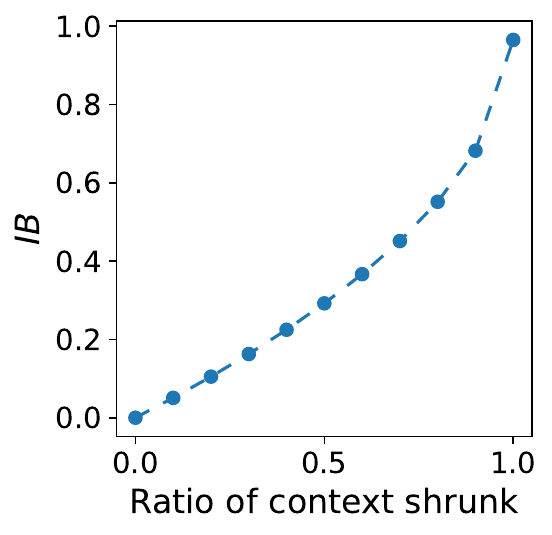}
        \caption{}
        \label{fig:shrunk_08_1000}
    \end{subfigure}
    \begin{subfigure}[b]{.32\linewidth}
        \includegraphics[width=\linewidth]{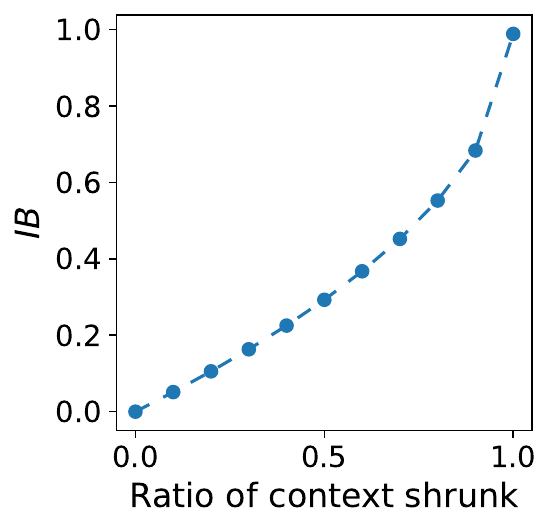}
        \caption{}
        \label{fig:shrunk_08_10000}
    \end{subfigure}
    \caption{$IB$ behaviour of the \textbf{shrinking} context of a node belonging to the minority community ((a), (b) and (c)) and majority community ((d), (e) and (f)). (a) and (d) correspond to graphs with 100 nodes, (b) and (e) correspond to graphs with 1,000 nodes, and (d) and (f) correspond to graphs with 10,000 nodes.}
    \label{fig:shrunk}
\end{figure}

Figure~\ref{fig:shrunk} shows the behaviour of $IB$ under context shrinking, where a node's predicted community is smaller than its ground-truth community. The node belongs to $c_m$ in plots~\ref{fig:shrunk_02_100},~\ref{fig:shrunk_02_1000}, and~\ref{fig:shrunk_02_10000} and to $c_M$ in plots~\ref{fig:shrunk_08_100},~\ref{fig:shrunk_08_1000}, and~\ref{fig:shrunk_08_10000}. 
It is noticeable that, despite the different community sizes of the minority and majority, the $IB$ scores for both nodes follow the same concave-up, increasing pattern, and their magnitudes are comparable, unlike the plots in Figure~\ref{fig:expanded}. Notably, the values for both $c_m$ and $c_M$ go up to $1$, indicating that the worst-case scenario under context shrinkage is more unfair than the worst-case scenario under context expansion according to $IB$.

Why does $IB$ underestimate context expansion as opposed to context shrinking? To compute $IB$, we use cosine similarity, which is inversely proportional to the product of the norms of the two vectors. Therefore, at minimum, this quantity will be equal to $1$ when $\|\Gamma_i\|=\|\Gamma'_i\|=1$, and that can only happen when both vectors have a singular non-zero value equal to $1$. Instead, regarding the dot product at the numerator, we have that $\Gamma_i \cdot \Gamma'_i\le\sum_{j=1}^{|V|}\Gamma_{i,j}$, because each entry in each vector can only be equal to $1$ or $0$, and our best case scenario is when $\Gamma_i=\Gamma'_i$. Consequently, since the ground-truth is fixed, expanding the context of the label will not increase the numerator, but it will increase the denominator. On the other hand, shrinking the context will decrease the numerator by $1$ for each entry changed and will decrease the denominator by a much smaller quantity. Due to this reason, adding nodes to a community produces less unfairness than removing nodes from it. This is an appreciated behaviour of $IB$, since adding nodes to a community preserves a node's original context, even though it may be altered. The same does not hold for removing nodes from the community.

\begin{figure}[t]
    \centering
    \begin{subfigure}[b]{.32\linewidth}
        \begin{center}
             100 nodes
        \end{center}
        \includegraphics[width=\linewidth]{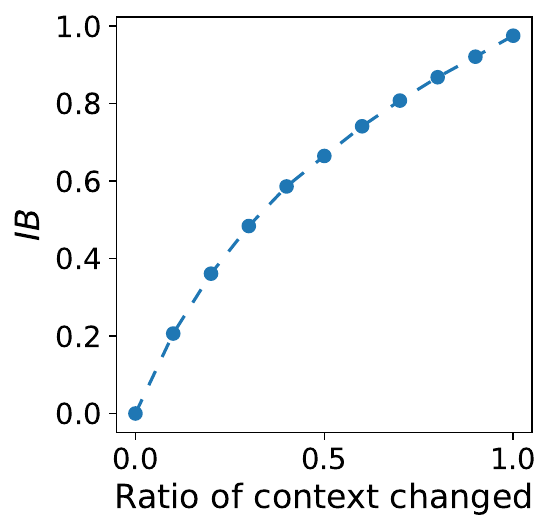}
        \caption{}
        \label{fig:changed_02_100}
    \end{subfigure}
    \begin{subfigure}[b]{.32\linewidth}
        \begin{center}
             1,000 nodes
        \end{center}
        \includegraphics[width=\linewidth]{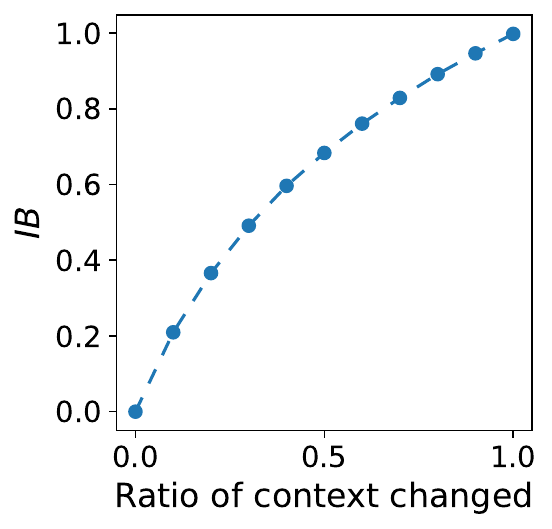}
        \caption{}
        \label{fig:changed_02_1000}
    \end{subfigure}
    \begin{subfigure}[b]{.32\linewidth}
        \begin{center}
             10,000 nodes
        \end{center}
        \includegraphics[width=\linewidth]{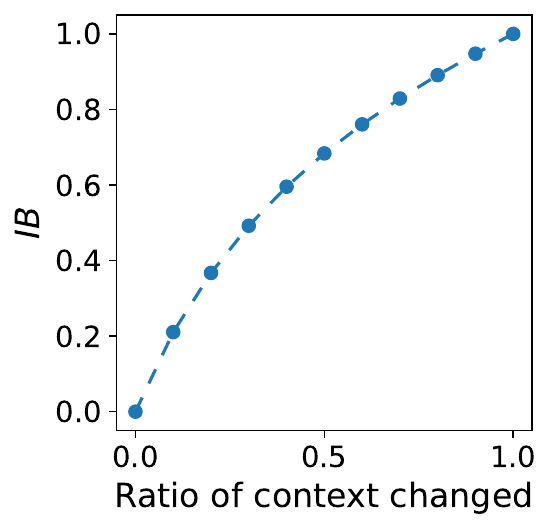}
        \caption{}
        \label{fig:changed_02_10000}
    \end{subfigure}
    \begin{subfigure}[b]{.32\linewidth}
        \includegraphics[width=\linewidth]{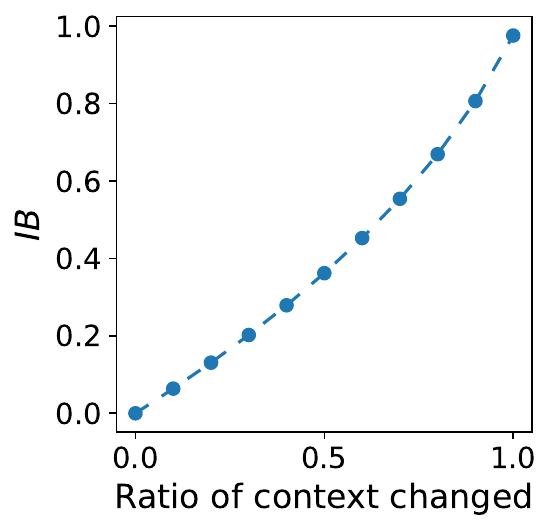}
        \caption{}
        \label{fig:changed_08_100}
    \end{subfigure}
    \begin{subfigure}[b]{.32\linewidth}
        \includegraphics[width=\linewidth]{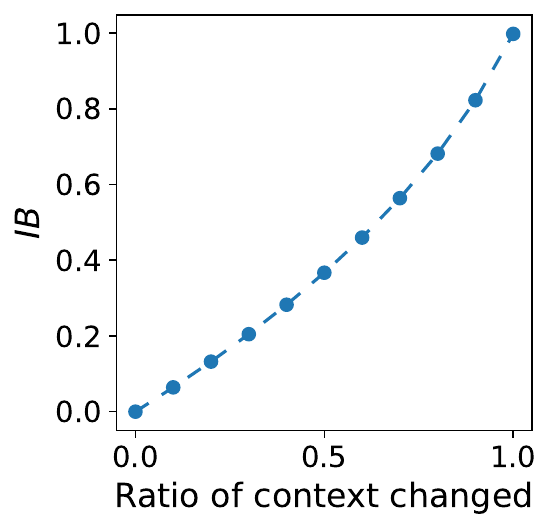}
        \caption{}
        \label{fig:changed_08_1000}
    \end{subfigure}
    \begin{subfigure}[b]{.32\linewidth}
        \includegraphics[width=\linewidth]{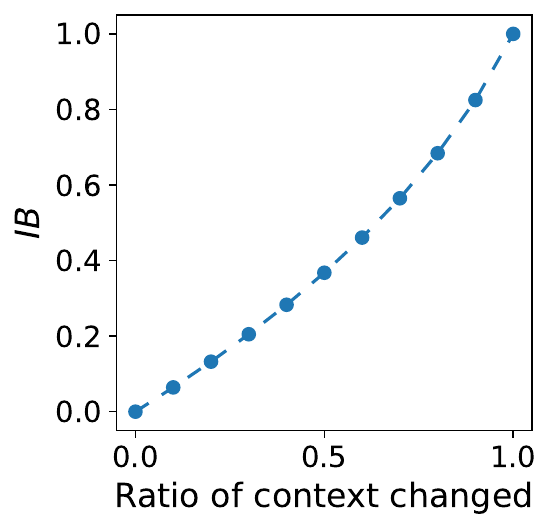}
        \caption{}
        \label{fig:changed_08_10000}
    \end{subfigure}
    \caption{$IB$ behaviour of the \textbf{changing} context of a node belonging to the minority community ((a), (b) and (c)) and majority community ((d), (e) and (f)). (a) and (d) correspond to graphs with 100 nodes, (b) and (e) correspond to graphs with 1,000 nodes, and (d) and (f) correspond to graphs with 10,000 nodes.}
    \label{fig:changed}
\end{figure}

Finally, Figure~\ref{fig:changed} shows the behaviour of $IB$ when a node's community is changed, which is effectively caused by a mix of context expansion and shrinkage. In simple terms, this means that the entries in the CC vector are swapped proportionally, from $0$ to $1$, and vice versa (except for the self-entry). When the ratio of nodes changed is $1.0$, the final result is $\Gamma'\approx\neg \Gamma$.

We observe that, for the minority community $c_m$, $IB$ behaves similarly to the expanding context (Figure~\ref{fig:expanded}), whereas for the majority community $c_M$, it resembles the shrinking context (Figure~\ref{fig:shrunk}). Changing the context of $c_m$ results in more nodes being added than removed, leading to a context alteration similar to a context expansion. On the other hand, changing the context of $c_M$ involves more node removals than additions, resulting in an alteration resembling a context shrinkage.

\begin{figure}
    \centering
    \begin{subfigure}[b]{.32\linewidth}
        \includegraphics[width=\linewidth]{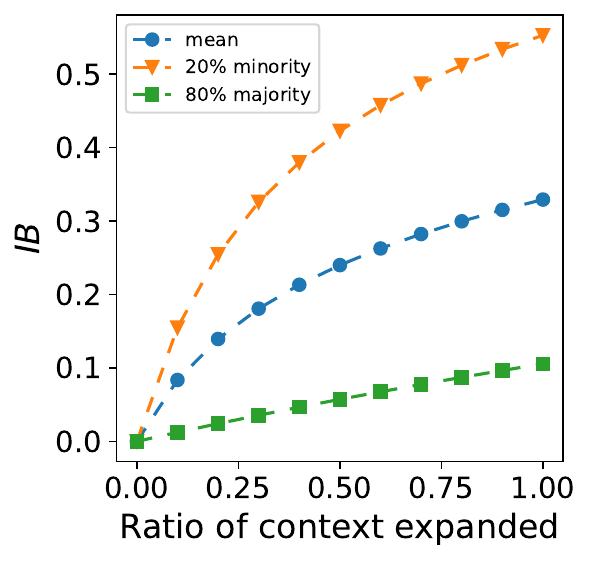}
        \caption{}
    \end{subfigure}
    \begin{subfigure}[b]{.32\linewidth}
        \includegraphics[width=\linewidth]{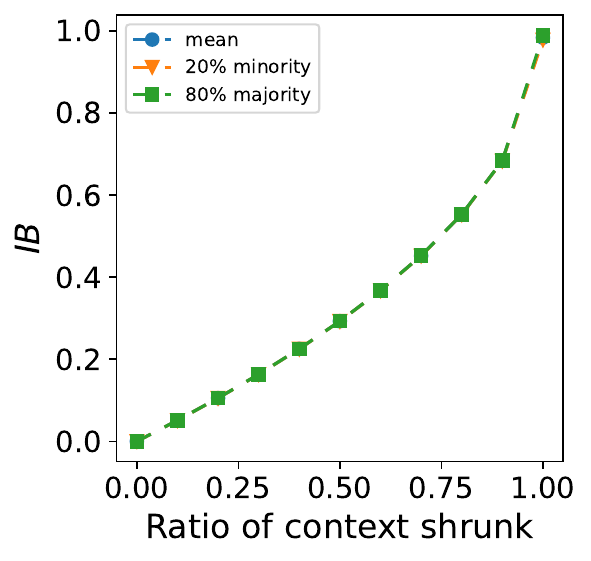}
        \caption{}
    \end{subfigure}
    \begin{subfigure}[b]{.32\linewidth}
        \includegraphics[width=\linewidth]{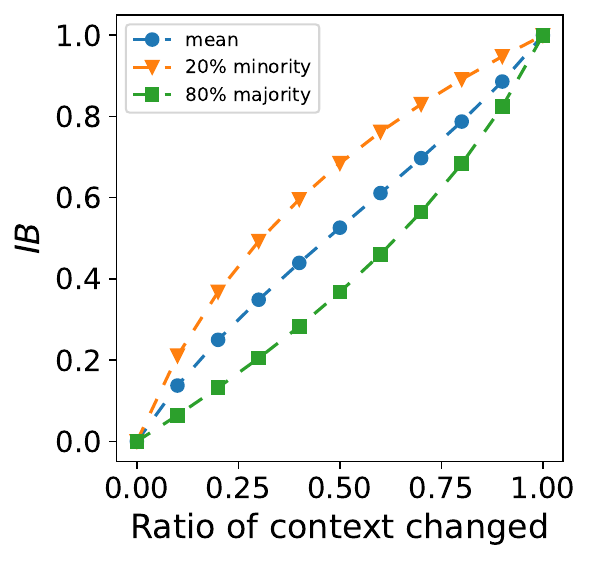}
        \caption{}
    \end{subfigure}
    \caption{Mean $IB$ for graph of 10,000 nodes in different context variation patterns: (a) context expansion, (b) context shrinking, and (c) context change. The blue line represents the mean of the minority community's behaviour (orange triangles) and the majority community's behaviour (green squares).}
    \label{fig:behaviour_mean}
\end{figure}

Figure~\ref{fig:behaviour_mean} shows the mean $IB$ values across the previous experiments. We present results on the 10,000-node graphs, as earlier plots demonstrated that graph size does not affect the analysis, and we observe similar behaviour for other sizes. In the expanded context subplot, the magnitude difference between $c_m$ and $c_M$ is clearly visible. In contrast, the shrinking context subplot shows the two curves overlapping almost perfectly. The changing context subplot exhibits a non-linear pattern: it is concave up between rates $0.0$ and $0.5$ and concave down between $0.5$ and $1.0$. This indicates that the context change curve for $c_m$ is steeper in the first half and smoother in the second, whereas the curve for $c_M$ shows the opposite behaviour.

\subsection{Computational Complexity}\label{subsec:computational_complexity}

In this subsection, we explain the computational complexity for the naive implementation of our method.

Given a network $G =(V, E)$ with $n=|V|$ nodes and two partitions of nodes into communities: the ground-truth $C=\{c_1,c_2,\ldots,c_n\}$ and the predicted communities $C'=\{c'_1,c'_2,\ldots,c'_n\}$, the Individual Bias ($IB$) is computed through two main stages:

\begin{enumerate}
    \item Construction of the \textbf{CC matrices} $\Gamma$ and $\Gamma'$.
    \item Computation of the \textbf{cosine distance} between the corresponding rows $\Gamma_i$ and $\Gamma'_i$ for all nodes $i\in V$.
\end{enumerate}

The computation of the CC matrix requires checking for every node pair $(i,j)$, yielding a time complexity of $O(n^2)$. Therefore, the time complexity to compute both $\Gamma$ and $\Gamma'$ is $O(n^2)$.

For each node $i$, the Individual Bias is computed as:
$$IB_i=1-\frac{\Gamma_i\cdot\Gamma'_i}{\|\Gamma_i\|\|\Gamma'_i\|}$$
Each dot product involves $n$ elements, resulting in $O(n^2)$ total operations for all nodes.
Therefore, the overall complexity to compute $IB$ for all nodes and graphs is $O(n^2)$.

\section{Experimental Setup}\label{sec:data_experimental_setup}

This section covers the datasets, CD algorithms, and evaluation measures used in our study. Our complete codebase is available in the corresponding GitHub repository\footnote{The repository will be made public with the accepted paper.}\footnote{\url{https://github.com/akratiiet/Individual_Fairness_in_Community_Detection}}.

\subsection{Datasets}
Experiments are performed on both synthetic and real-world networks as explained below.

\subsubsection{Synthetic dataset}\label{subsec:synth_dataset}
The synthetic networks were generated using the ABCD benchmark model, introduced by Kamiński et al.~\cite{Kami_ski_2023} as an improvement over the classical LFR benchmark~\cite{Lancichinetti_2008}. ABCD produces networks with a built-in community structure and thus provides ground-truth partitions for evaluation. However, ABCD offers several advantages over LFR: it is faster, scalable to larger graphs, and it can be conceptualized as a union of independent random graphs, which facilitates theoretical analysis using existing tools from random graph theory.

Next, we explain the parameters used in the ABCD benchmark model, and their values used in the network generation are mentioned in bold.
\begin{itemize}
    \item \textbf{n}: Number of vertices in the generated graph \textbf{(n =10,000)}
    \item \textbf{$\gamma$}: Power-law exponent for degree distribution \textbf{($\gamma$ = 2.5)}
    \item \textbf{d\_min}: Minimum degree \textbf{(d\_min = 5)}
    \item \textbf{d\_max}: Maximum degree \textbf{(d\_max=50)}
    \item \textbf{d\_max\_iter}: Maximum number of iterations for sampling degrees \textbf{(d\_max\_iter = 1,000)}
    \item \textbf{$\beta$}: Power-law exponent for cluster size distribution \textbf{($\beta$ = 1.5)}
    \item \textbf{c\_min}: Minimum cluster size \textbf{(c\_min = 100)}
    \item \textbf{c\_max}: Maximum cluster size \textbf{(c\_max = 1,000)}
    \item \textbf{c\_max\_iter}: Maximum number of iterations for sampling cluster sizes \textbf{(c\_max\_iter = 1,000)}
    \item \textbf{$\xi$}: Fraction of edges placed as the inter-community connections (or in the background graph). It offers a more natural interpretation than LFR's $\mu$ parameter, avoiding the ``anti-community" effect observed in LFR with high mixing parameters ($\mu$). A $\xi$ value of $0$ means all edges are within communities, while a $\xi$ value of $1$ results in a pure random graph with no community effect on edge generation. \textbf{($\xi$~=~0.2, 0.4, 0.6)}
\end{itemize}

The $\xi$ parameter is of special interest to us, as varying it allows us to study how $IB$ and $IB_G$ vary according to the ``explicitness" of the community structure, with $\xi = 0.2$ producing very clear communities and $\xi = 0.6$ creating highly obscure ones. 
We generate three datasets corresponding to three values of $\xi$, with 10 networks per value. For each dataset, all CD methods are run on all 10 networks, and we report the mean and standard deviation of the results.

\subsubsection{Real-world Networks}\label{subsubsec:real_world_networks}
We perform experiments on the following three real-world networks. The properties of these networks are listed in Table~\ref{tab:real_world_data}.

\begin{table}[t]
    \centering
    \begin{tabular}{l c c c}
        \toprule
        \textbf{Dataset} & $|V|$ & $|E|$ & $|C|$ \\
        \midrule
        Email-Eu-core & 1005 & 25571 & 42 \\
        Football & 115 & 613 & 12 \\
        Polbooks & 105 & 441 & 3 \\
        \bottomrule
    \end{tabular}
    \caption{Statistics about the real-world networks we used}
    \label{tab:real_world_data}
\end{table}

\begin{itemize}
    \item Email-Eu-core~\cite{yin2017local, leskovec2007graph}: This dataset is generated from the internal email communication of a European research organization. The ground-truth communities correspond to the department of the research institute to which the nodes belong. The finally generated network is undirected.
    \item Football~\cite{girvan2002community}: This network is obtained from the schedule of a football division of USA colleges, in the year 2000. Vertices represent teams, and edges represent games between the teams they connect. The teams are divided into ``conferences", and teams from the same conference are more likely to play against each other. These conferences are our ground-truth communities.
    \item Polbooks~\cite{orgnet}: The nodes in this network represent books about USA politics sold by the online bookseller Amazon.com. Edges represent frequent co-purchasing of books by the same buyers, as indicated by the ``customers who bought this book also bought these other books" feature on Amazon. The ground-truth communities are the political leaning of the book: liberal, neutral, or conservative.
\end{itemize}

\subsection{Community detection algorithms}\label{subsec:cd_algs}
In our experiments, we evaluate 30 CD algorithms, which are grouped into seven categories based on their approaches as mentioned below. Table~\ref{tab:cd_algs} lists all these algorithms grouped by category.

\begin{itemize}
    \item \textbf{Optimisation}: Communities are identified by maximising an objective function that evaluates partition structure. Most approaches optimise modularity~\cite{newman2004finding}, with the exception of Significance~\cite{traag2013significant}. Because modularity optimisation is NP-hard, these methods rely on heuristic strategies.
    
    \item \textbf{Hierarchical Clustering/Agglomerative}: Hierarchical methods recursively partition a network or merge groups to create a tree structure, called a dendrogram, which shows communities at different scales. Walktrap~\cite{pons2005computing}, for example, iteratively merges the closest communities, often measuring similarity based on random walks.
    
    \item \textbf{Spectral}: These methods partition the network using the spectral properties of matrices such as the adjacency or Laplacian matrices~\cite{fortunato2016community}. These approaches exploit eigenvalues and eigenvectors to reveal structural patterns. Spectral Clustering~\cite{Higham2007}, for instance, uses the Fiedler vector (the eigenvector associated with the second smallest Laplacian eigenvalue) to form communities.
    
    \item \textbf{Information Flow}: This category identifies community structure by modelling the flow of information or dynamics across the network. It typically uses the probability flow of random walks as a proxy for how easily information moves between nodes. The Infomap~\cite{Rosvall_2008} algorithm, for instance, minimizes the expected description code length of a random walk to determine the optimal modular structure.
    
    \item \textbf{Statistical}: Statistical methods approach CD as an inference problem, aiming to discover the most likely partition that explains the observed network connections. They rely on fitting the network to a generative model, such as the Stochastic Block Model (SBM)~\cite{HOLLAND1983109}. The fitting process often utilizes maximum likelihood estimation, sometimes employing complex iterative techniques like the expectation-maximization (EM)~\cite{newman2007mixture} algorithm to infer hidden group assignments. This approach allows for determining definitions of groups and detecting specialized structures beyond simple assortative mixing
    
    \item \textbf{Network Embedding}: These methods first learn a low-dimensional embedding of the network and then apply clustering algorithms (e.g., k-means) to obtain the final partition~\cite{arya2022node}. The embedding captures structural similarities, allowing clustering methods to identify communities.

    \item \textbf{Others}: This category includes other algorithms that do not fall in the previous categories.
\end{itemize}

\begin{table}[t]
\centering
\footnotesize
\begin{tabular}{l l}
\toprule
\textbf{CD Algorithm} & \textbf{Approach} \\
\midrule
\multicolumn{2}{c}{\textbf{1. Optimisation}} \\
\midrule
Louvain~\cite{Blondel_2008} & Hierarchical greedy modularity optimization \\
Leiden~\cite{Traag_2019} & Improved Louvain with better partition refinement \\
CNM~\cite{clauset2004finding} & Greedy agglomerative modularity maximization \\
CPM~\cite{traag2011narrow} & Optimizes Potts model with resolution parameter \\
Rb pots~\cite{reichardt2006statistical, leicht2008community} & Finds the ground state of an infinite range spin glass \\
Rber pots~\cite{reichardt2006statistical} & Variant of Rb pots with ER random null model \\
Spinglass~\cite{reichardt2006statistical} & Statistical physics Potts model \\
Combo~\cite{Sobolevsky_2014} & Universal optimiser \\
Significance~\cite{traag2013significant} & Evaluates partition Significance based on subgraph probabilities \\
\midrule
\multicolumn{2}{c}{\textbf{2. Embedding}} \\
\midrule
DeepWalk~\cite{perozzi2014deepwalk} & Learns node embeddings from truncated random walks \\
Node2vec~\cite{grover2016node2vec} & Biased random-walk node embedding \\
DER~\cite{kozdoba2015community} & Embeds nodes using random-walk measures with k-means clustering \\
\midrule
\multicolumn{2}{c}{\textbf{3. Hierarchical Clustering/Agglomerative}} \\
\midrule
Walktrap~\cite{pons2005computing} & Hierarchical clustering via random walk similarity \\
AGDL~\cite{zhang2012graph} & Agglomerative clustering using indegree/outdegree product \\
Paris~\cite{bonald2018hierarchical} & Hierarchical agglomerative clustering using dendogram \\
\midrule
\multicolumn{2}{c}{\textbf{4. Spectral}} \\
\midrule
Spectral~\cite{HIGHAM200725} & Graph embedding using eigenvectors \\
Eigenvector~\cite{newman2006finding} & Splits network based on modularity matrix eigenvector \\
Kcut~\cite{ruan2007efficient} & Recursive k-way partitioning \\
\midrule
\multicolumn{2}{c}{\textbf{5. Information Flow}} \\
\midrule
Infomap~\cite{Rosvall_2008} & Uses random walks to approximate information flow \\
Fluid communities~\cite{pares2017fluid} & Label propagation with fluid dynamics \\
Label propagation~\cite{label_propagation} & Node label consensus from neighbors \\
Chinese Whispers~\cite{chinese_whispers} & Randomized graph-embedding, seeking neighborhood consensus \\
Markov clustering~\cite{enright2002efficient} & Simulates random walks with inflation/expansion \\
\midrule
\multicolumn{2}{c}{\textbf{6. Statistical}} \\
\midrule
SBMDL~\cite{peixoto2014efficient} & Infers modular structure minimizing description length \\
EM~\cite{newman2007mixture} & Classifies vertices probabilistically into groups \\
Surprise~\cite{Traag_2015} & Optimizes a measure called surprise, to assess the quality of partitions \\
Belief Propagation~\cite{zhang2014scalable} & Approximate marginals of a Gibbs distribution, seeking consensus \\
Head Tail~\cite{Jiang_2015} & Iteratively partitions network via edge betweenness breaks \\
\midrule
\multicolumn{2}{c}{\textbf{7. Others}} \\
\midrule
Ricci Flow~\cite{ni2019community} & Applies discrete Ricci flow to edge weights \\
MCODE~\cite{BaderGaryD2003Aamf} & Detects dense regions in protein networks \\
\bottomrule
\end{tabular}
\caption{Algorithms used for the experiments: they are categorised in seven different groups based on their approaches.}
\label{tab:cd_algs}
\end{table}

\subsection{Community Quality Measures}\label{subsec:cd_scores}

To evaluate the structural quality of the detected communities, we use four widely used measures (see Table~\ref{tab:cd_scores}). These were chosen to complement our fairness measures and to analyse the performance–fairness trade-offs. The measures fall into two categories:

\begin{itemize}
    \item \textbf{Internal}: Internal measures completely rely on the predicted partition to evaluate the quality of the identified communities. We use Modularity~\cite{newman2004finding} as the representative internal measure, given its popularity in the literature.
    \item \textbf{External}: External measures compare predicted communities against ground-truth labels. We use Normalized Mutual Information (NMI)~\cite{lancichinetti2009detecting}, Adjusted Rand Index (ARI)~\cite{ari}, and Normalized F1 (NF1)~\cite{nf1, nf1_2} to capture different perspectives of similarity.
\end{itemize}

\begin{table}[t]
    \centering
    \begin{tabular}{p{4cm}p{4cm} c}
        \toprule
        \textbf{Measure} & \textbf{Description} & \textbf{Type} \\
        \midrule
        Newman Girvan Modularity~\cite{newman2004finding} & The normalised difference between the number of edges in a community and the expected number of edges in it according to a null model. & Internal \\
        \midrule
        Normalized Mutual Information (NMI)~\cite{lancichinetti2009detecting} & Normalization of the Mutual Information score, calculated as $S(c1) + S(c2) - S(c1, c2)$, where $S$ is the Shannon Entropy. & External \\
        \midrule
        Adjusted Rand Index (ARI)~\cite{ari} & The Rand Index ($RI$), that computes a similarity measure between two clusterings by considering all pairs of samples and counting pairs, adjusted for chance. & External \\
        \midrule
        Normalized F1 (NF1)~\cite{nf1, nf1_2} & Normalization of the F1 score applied to communities. & External \\
        \bottomrule
    \end{tabular}
    \caption{Community quality measures, with a brief description and their affiliation to a score family.}
    \label{tab:cd_scores}
\end{table}

\subsection{Group fairness measure}\label{subsec:group_fairness_measure}
The individual fairness results will be compared with the group fairness measure $\Phi$~\cite{devink2024groupfairnessmetricscommunity}.
The measure $\Phi$ (mentioned in Section~\ref{sec:related_work}) computes the group fairness of a CD algorithm, putting in relation predicted communities' quality scores with corresponding ground-truth communities' properties. After fitting a line between these points, $\Phi$ is obtained by the angular coefficient of said line. The community properties taken in consideration are size, conductance, and density. The community quality measures are Fraction of Correctly Classified Nodes (FCCN), F1 score, and Fraction of Correctly Classified Edges (FCCE).

\section{Results}\label{sec:results}

The results section is organized into five subsections: individual bias analysis, the trade-off between Individual Bias and performance, a comparison between individual and group fairness, an analysis of individual bias distributions, and results on real-world networks.

\subsection{Individual Bias Analysis}\label{subsec:ibg_results_analysis}

\begin{table}[t]
    \centering
    \resizebox{\textwidth}{!}{
    \begin{tabular}{lccc}
    \toprule
    {} & $\xi=0.2$ & $\xi=0.4$ & $\xi=0.6$ \\
    \midrule
    Louvain & $1.64\mathrm{e}{-02}\pm2.82\mathrm{e}{-02}$ & $2.01\mathrm{e}{-01}\pm2.28\mathrm{e}{-02}$ & $8.38\mathrm{e}{-02}\pm1.40\mathrm{e}{-02}$ \\
    Leiden & $1.22\mathrm{e}{-16}\pm1.28\mathrm{e}{-17}$ & $7.65\mathrm{e}{-02}\pm3.01\mathrm{e}{-02}$ & $8.91\mathrm{e}{-02}\pm1.71\mathrm{e}{-02}$ \\
    CNM & $2.17\mathrm{e}{-01}\pm1.42\mathrm{e}{-02}$ & $1.78\mathrm{e}{-01}\pm1.44\mathrm{e}{-02}$ & $5.63\mathrm{e}{-02}\pm4.00\mathrm{e}{-03}$ \\
    CPM & $1.22\mathrm{e}{-16}\pm1.28\mathrm{e}{-17}$ & $3.49\mathrm{e}{-01}\pm3.15\mathrm{e}{-02}$ & $7.92\mathrm{e}{-02}\pm2.07\mathrm{e}{-02}$ \\
    Rb pots & $1.22\mathrm{e}{-16}\pm1.28\mathrm{e}{-17}$ & $9.62\mathrm{e}{-02}\pm2.81\mathrm{e}{-02}$ & $8.19\mathrm{e}{-02}\pm8.73\mathrm{e}{-03}$ \\
    Rber pots & $1.22\mathrm{e}{-16}\pm1.28\mathrm{e}{-17}$ & $3.29\mathrm{e}{-01}\pm7.42\mathrm{e}{-02}$ & $7.90\mathrm{e}{-02}\pm2.08\mathrm{e}{-02}$ \\
    Spinglass & $1.52\mathrm{e}{-01}\pm1.62\mathrm{e}{-02}$ & $1.57\mathrm{e}{-01}\pm1.86\mathrm{e}{-02}$ & $2.94\mathrm{e}{-01}\pm7.06\mathrm{e}{-03}$ \\
    Combo & $1.22\mathrm{e}{-16}\pm1.28\mathrm{e}{-17}$ & $1.72\mathrm{e}{-02}\pm1.56\mathrm{e}{-02}$ & $6.71\mathrm{e}{-02}\pm1.20\mathrm{e}{-02}$ \\
    Significance & $3.90\mathrm{e}{-01}\pm1.19\mathrm{e}{-02}$ & $3.02\mathrm{e}{-01}\pm2.22\mathrm{e}{-02}$ & $8.11\mathrm{e}{-02}\pm1.32\mathrm{e}{-02}$ \\
    Deepwalk & $1.93\mathrm{e}{-01}\pm1.89\mathrm{e}{-02}$ & $2.33\mathrm{e}{-01}\pm1.61\mathrm{e}{-02}$ & $2.44\mathrm{e}{-01}\pm2.32\mathrm{e}{-02}$ \\
    Node2vec & $1.99\mathrm{e}{-01}\pm2.70\mathrm{e}{-02}$ & $2.25\mathrm{e}{-01}\pm2.59\mathrm{e}{-02}$ & $2.46\mathrm{e}{-01}\pm2.19\mathrm{e}{-02}$ \\
    DER & $8.76\mathrm{e}{-02}\pm9.38\mathrm{e}{-03}$ & $9.61\mathrm{e}{-02}\pm8.14\mathrm{e}{-03}$ & $5.39\mathrm{e}{-02}\pm5.10\mathrm{e}{-03}$ \\
    Walktrap & $1.52\mathrm{e}{-01}\pm1.05\mathrm{e}{-02}$ & $3.13\mathrm{e}{-01}\pm5.83\mathrm{e}{-03}$ & $5.50\mathrm{e}{-02}\pm7.35\mathrm{e}{-03}$ \\
    AGDL & $1.04\mathrm{e}{-01}\pm1.93\mathrm{e}{-02}$ & $6.95\mathrm{e}{-02}\pm8.97\mathrm{e}{-03}$ & $5.94\mathrm{e}{-02}\pm1.07\mathrm{e}{-02}$ \\
    Paris & $1.97\mathrm{e}{-02}\pm2.67\mathrm{e}{-03}$ & $1.35\mathrm{e}{-02}\pm7.58\mathrm{e}{-04}$ & $1.27\mathrm{e}{-02}\pm4.35\mathrm{e}{-04}$ \\
    Spectral & $8.94\mathrm{e}{-02}\pm9.17\mathrm{e}{-03}$ & $9.21\mathrm{e}{-02}\pm1.09\mathrm{e}{-02}$ & $5.89\mathrm{e}{-02}\pm6.07\mathrm{e}{-03}$ \\
    Eigenvector & $2.27\mathrm{e}{-01}\pm3.40\mathrm{e}{-02}$ & $1.15\mathrm{e}{-01}\pm1.36\mathrm{e}{-02}$ & $4.64\mathrm{e}{-02}\pm6.28\mathrm{e}{-03}$ \\
    Kcut & $6.15\mathrm{e}{-02}\pm6.27\mathrm{e}{-03}$ & $6.15\mathrm{e}{-02}\pm6.30\mathrm{e}{-03}$ & $6.15\mathrm{e}{-02}\pm6.29\mathrm{e}{-03}$ \\
    Infomap & $2.59\mathrm{e}{-02}\pm7.15\mathrm{e}{-02}$ & $3.40\mathrm{e}{-01}\pm5.00\mathrm{e}{-02}$ & $1.04\mathrm{e}{-01}\pm2.02\mathrm{e}{-02}$ \\
    Fluid communities & $2.78\mathrm{e}{-01}\pm9.86\mathrm{e}{-03}$ & $1.69\mathrm{e}{-01}\pm3.22\mathrm{e}{-02}$ & $3.16\mathrm{e}{-02}\pm3.13\mathrm{e}{-03}$ \\
    Label propagation & $1.10\mathrm{e}{-01}\pm3.85\mathrm{e}{-02}$ & $2.32\mathrm{e}{-01}\pm7.18\mathrm{e}{-02}$ & $6.14\mathrm{e}{-02}\pm6.27\mathrm{e}{-03}$ \\
    Chinese Whispers & $1.22\mathrm{e}{-16}\pm1.28\mathrm{e}{-17}$ & $2.89\mathrm{e}{-01}\pm7.37\mathrm{e}{-02}$ & $6.15\mathrm{e}{-02}\pm6.28\mathrm{e}{-03}$ \\
    Markov clustering & $3.32\mathrm{e}{-01}\pm1.13\mathrm{e}{-02}$ & $2.18\mathrm{e}{-01}\pm3.29\mathrm{e}{-02}$ & $6.88\mathrm{e}{-02}\pm6.88\mathrm{e}{-03}$ \\
    SBMDL & $5.14\mathrm{e}{-03}\pm1.54\mathrm{e}{-02}$ & $7.43\mathrm{e}{-02}\pm7.75\mathrm{e}{-02}$ & $6.43\mathrm{e}{-02}\pm9.48\mathrm{e}{-03}$ \\
    EM & $2.76\mathrm{e}{-01}\pm9.89\mathrm{e}{-03}$ & $8.92\mathrm{e}{-02}\pm7.31\mathrm{e}{-03}$ & $2.04\mathrm{e}{-02}\pm1.08\mathrm{e}{-03}$ \\
    Surprise & $2.44\mathrm{e}{-01}\pm4.94\mathrm{e}{-02}$ & $1.94\mathrm{e}{-01}\pm2.94\mathrm{e}{-02}$ & $5.72\mathrm{e}{-02}\pm5.43\mathrm{e}{-03}$ \\
    Belief propagation & $1.48\mathrm{e}{-01}\pm1.96\mathrm{e}{-02}$ & $1.67\mathrm{e}{-01}\pm1.89\mathrm{e}{-02}$ & $1.84\mathrm{e}{-01}\pm1.73\mathrm{e}{-02}$ \\
    Head Tail & $2.76\mathrm{e}{-02}\pm1.00\mathrm{e}{-03}$ & $2.94\mathrm{e}{-02}\pm1.03\mathrm{e}{-03}$ & $2.38\mathrm{e}{-02}\pm8.45\mathrm{e}{-04}$ \\
    Ricci flow & $2.34\mathrm{e}{-01}\pm1.20\mathrm{e}{-01}$ & $5.75\mathrm{e}{-02}\pm6.68\mathrm{e}{-03}$ & --- \\
    MCODE & $2.94\mathrm{e}{-02}\pm2.60\mathrm{e}{-03}$ & $2.14\mathrm{e}{-02}\pm9.51\mathrm{e}{-04}$ & $1.93\mathrm{e}{-02}\pm1.03\mathrm{e}{-03}$ \\
    \bottomrule
    \end{tabular}}
    \caption{Experimental mean values of $IB_G$ for different $\xi$ values.}
    \label{tab:IBg_results}
\end{table}

Table~\ref{tab:IBg_results} shows the $IB_G$ values, averaged over the ten ABCD graphs for each $\xi$, along with their standard deviations. Most methods show a small standard deviation, indicating highly consistent graph-level individual bias behaviour across networks. The most notable exception is Ricci Flow for $\xi = 0.2$, with a standard deviation of $0.12$. The main reason for this standard deviation value is related to the ABCD graph structure: Ricci Flow should assign positive curvature values to intra-community edges, and negative values to inter-community edges, and then eliminate the negative ones to isolate communities. However, because the generated ABCD graphs are relatively sparse, the method tends to assign positive curvature to edges incident to high-degree nodes. This leads to the formation of a large cluster (roughly between 5,000 and 9,000 nodes) containing the main hubs of the networks, while the remaining nodes are split into many small communities.

For networks with highly explicit community structure ($\xi = 0.2$), the Significance CD method has the highest $IB_G$, indicating strong individual unfairness. In contrast, Leiden, CPM, Rb Pots, Rber Pots, Combo, and Chinese Whispers achieve $IB_G$ values on the order of $10^{-16}$, effectively zero. These algorithms, therefore, treat all nodes equally from an individual fairness perspective. Other methods with low $IB_G$ at $\xi = 0.2$ include SBMDL, Louvain, Infomap, Head Tail, MCODE, and Paris.

When $\xi = 0.4$, the fairest algorithms are Combo, Paris, Head Tail, and MCODE, each achieving $IB_G < 0.03$. The least fair method is CPM, with $IB_G = 0.35$, though this remains close to the maximum unfairness observed at $\xi = 0.2$ (Significance Communities with $IB_G = 0.39$). Notably, CPM shifts from being one of the fairest methods at $\xi = 0.2$ to one of the least fair at $\xi = 0.4$. The reason for this sudden change, in terms of fairness, is bound to an expected drop in performance as communities become harder to detect. This leads to discrimination towards certain nodes in favour of others.

For networks with obscure community structure ($\xi = 0.6$), Paris, Head Tail, and MCODE again appear among the fairest algorithms, along with the Eigenvector, Fluid Communities, and EM methods, all with $IB_G < 0.05$. Most methods have similarly low $IB_G$ in these networks, and the reason is that, since communities are extremely hard to detect, almost all nodes are treated equally unfairly. Therefore, the algorithms are being fair towards the graph, not favouring any particular group of nodes. Exceptions include Spinglass, DeepWalk, and Node2vec, each showing $IB_G > 0.2$. The reason being that these methods have a wider spread of individual fairness values on the node level at $\xi=0.6$. This means a better performance at the cost of fairness, since nodes get treated differently. It is worth noting that the Ricci community algorithm consistently failed to detect communities for $\xi=0.6$. The error returned was ``No cutoff point found!", meaning that the algorithm can not clearly find edges that bridge between different clusters. 
Overall, across $\xi = 0.4$ and $\xi = 0.6$, the $IB_G$ values are very small, as communities become harder to detect. In the extreme case where communities are effectively undetectable and assignments approach randomness, all nodes are treated similarly, leading to $IB_G \approx 0$. This observation highlights a significant limitation of individual fairness metrics: when detection quality deteriorates, perfect fairness may become meaningless, as uniformly poor treatment will be completely fair under $IB_G$.

\subsection{Individual Bias vs. Performance Trade-off}\label{subsec:ibg_results}

\begin{figure}[t!]
    \centering
    \begin{subfigure}[b]{\linewidth}
        \centering
        \includegraphics[width=.83\linewidth]{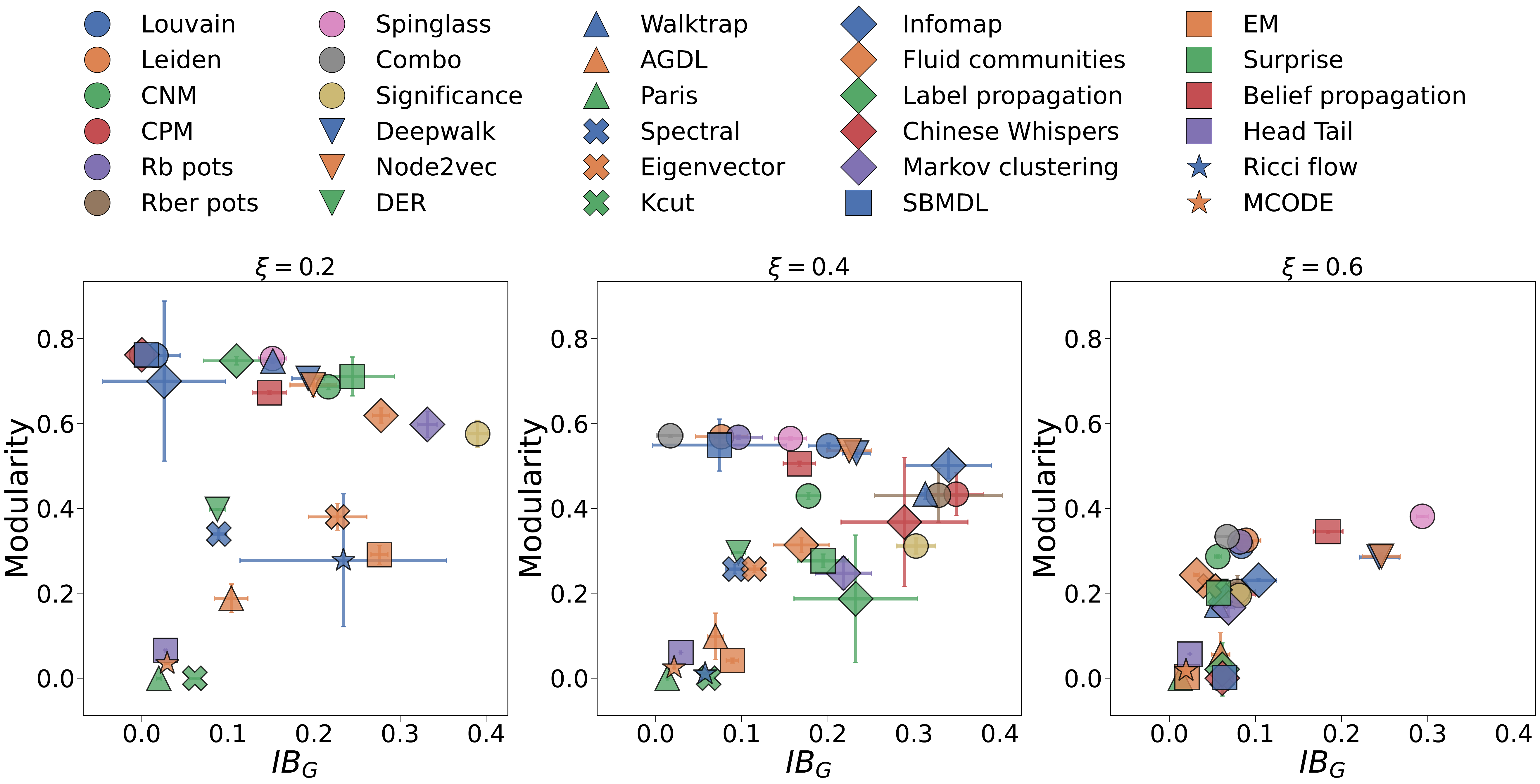}
        \label{fig:IBg_vs_Modularity}
    \end{subfigure}
    \begin{subfigure}[b]{\linewidth}
        \centering
        \includegraphics[trim={0 0 0 14cm}, clip, width=.83\linewidth]{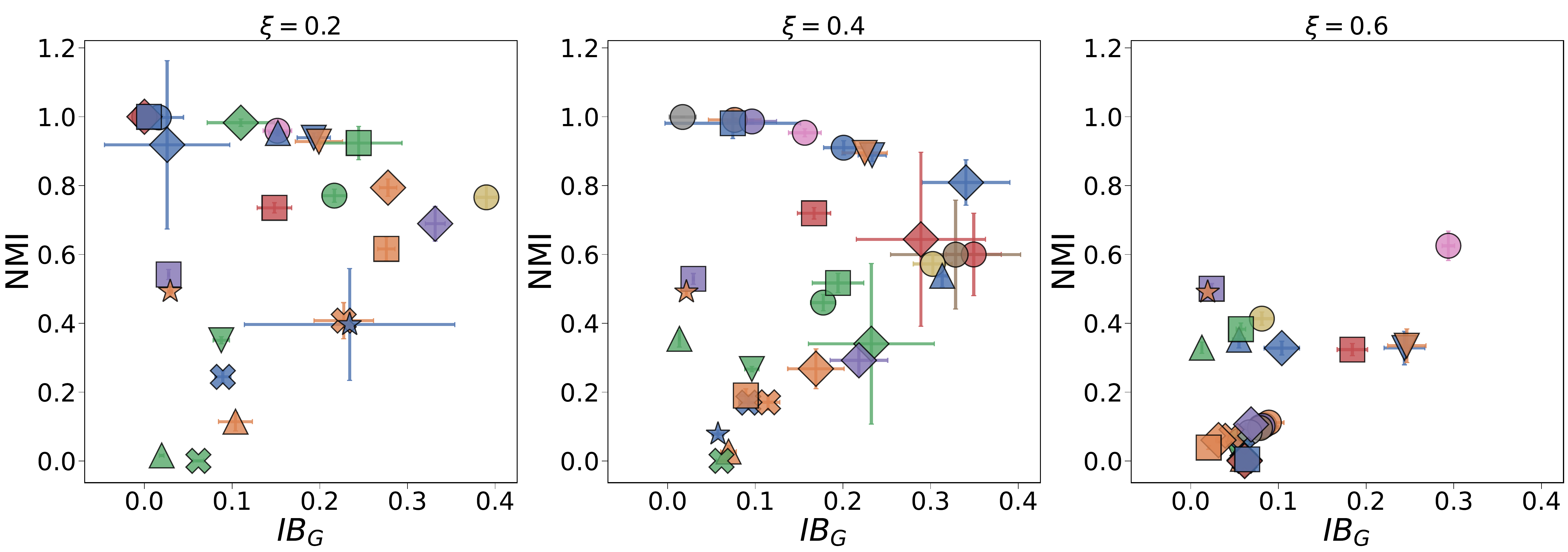}
        \label{fig:IBg_vs_NMI}
    \end{subfigure}
    \begin{subfigure}[b]{\linewidth}
        \centering
        \includegraphics[trim={0 0 0 14cm}, clip, width=.83\linewidth]{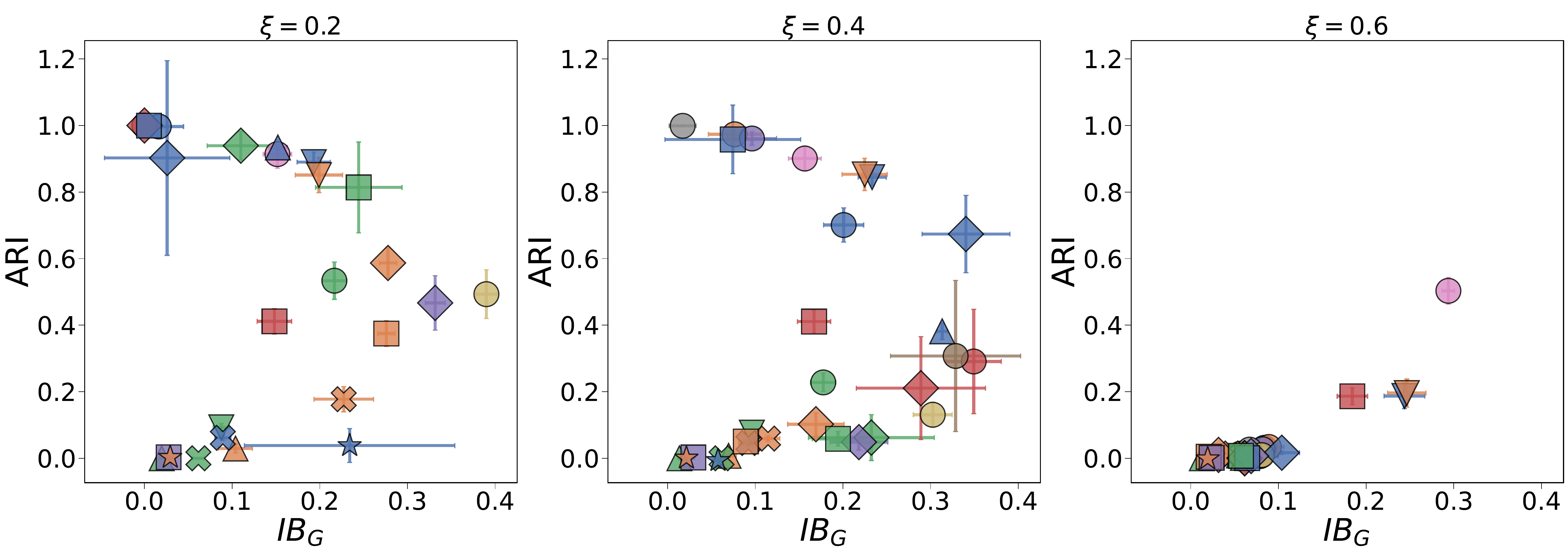}
        \label{fig:IBg_vs_ARI}
    \end{subfigure}
    \begin{subfigure}[b]{\linewidth}
        \centering
        \includegraphics[trim={0 0 0 14cm}, clip, width=.83\linewidth]{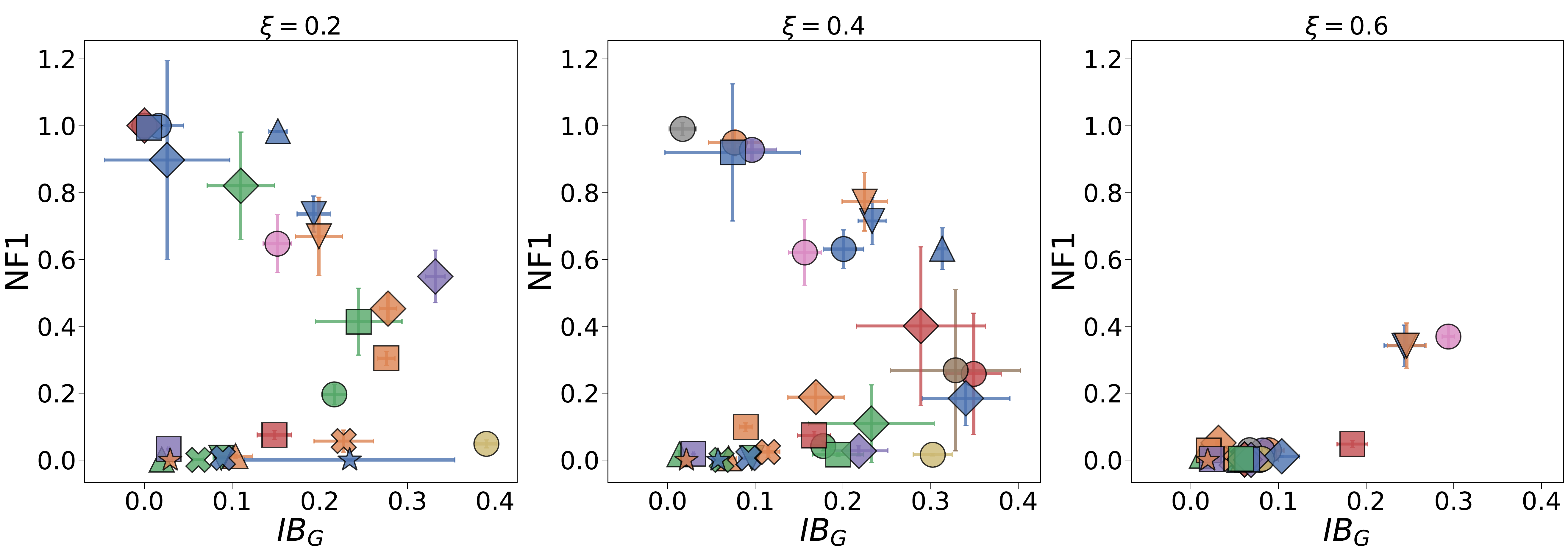}
        \label{fig:IBg_vs_NF1}
    \end{subfigure}
    \caption{$IB_G$ measure against communities' quality scores for $\xi=0.2$, $\xi=0.4$ and $\xi=0.6$. Each value is a mean value obtained from the analysis on 10 different graphs. The error lines are the standard deviations of these values}
    \label{fig:IBg_vs_performance}
\end{figure}

To further enrich our analysis, we plot the individual bias against the community quality metrics defined in Section~\ref{subsec:cd_scores}. Figure~\ref{fig:IBg_vs_performance} shows $IB_G$ values versus Modularity, NMI, ARI, and NF1, respectively. All values represent means over ten graphs per $\xi$, with standard deviation indicated by error bars. 

In Figure~\ref{fig:IBg_vs_performance}, modularity tends to overestimate the performance of optimisation-based methods, particularly for $\xi=0.6$, which is expected given that many of these algorithms directly optimise modularity.

ARI, shown in Figure~\ref{fig:IBg_vs_performance}, is more discriminative: algorithms tend to cluster either at the top or bottom of the plot, with few mid-performing methods.
This polarisation becomes stronger for $\xi=0.4$.

In Figure~\ref{fig:IBg_vs_performance}, NF1 similarly highlights clustering CD algorithms in high-performing and low-performing algorithms. NF1 tends to underestimate performance for methods that overestimate the number of communities. Significance Communities and Spinglass illustrate this effect clearly at $\xi=0.2$ and $\xi=0.4$. At $\xi=0.6$, almost all algorithms score near $0$, with only Spinglass, DeepWalk, and Node2vec reaching $NF1\approx 0.4$. This relates to the same reasoning that we explained in the previous subsection regarding these three algorithms $IB_G$ values.

An interesting trend is observable, especially for $\xi=0.2$ and $\xi=0.4$: to better explain it, we will refer to the top part of the plots as the ``best performing algorithms", and to the bottom part of the plots as the ``worst performing algorithms". In the top part (``best-performing region”), performance decreases as $IB_G$ increases. In contrast, at the bottom part (``worst-performing region”), performance improves as $IB_G$ increases. In other words, the fairest algorithms are those that either perform extremely well or extremely poorly, while the least fair algorithms tend to lie in the middle. For $\xi=0.6$, the best-performing region is mostly unpopulated.
Finally, these results show that no class of CD algorithms is inherently fairer than others. Fairness and performance vary widely based on network structure, highlighting that no single algorithm can be considered universally optimal. For example, Spinglass and the embedding methods outperform other methods when communities are very obscure. When communities are very explicit, Infomap, Chinese Whispers, SBMDL, Louvain, Leiden, Label Propagation, Walktrap, Spinglass, Surprise, CPM, Rb pots, Rber pots, Deepwalk, and Node2Vec are among the algorithms that have a perfect performance/fairness trade-off. We also noticed that Combo is the most resilient algorithm both in terms of fairness and community quality when communities become less explicit.

\subsection{Individual Fairness vs. Group Fairness}\label{subsec:individual_vs_group}
\begin{figure}[!t]
    \centering
    \includegraphics[width=\linewidth]{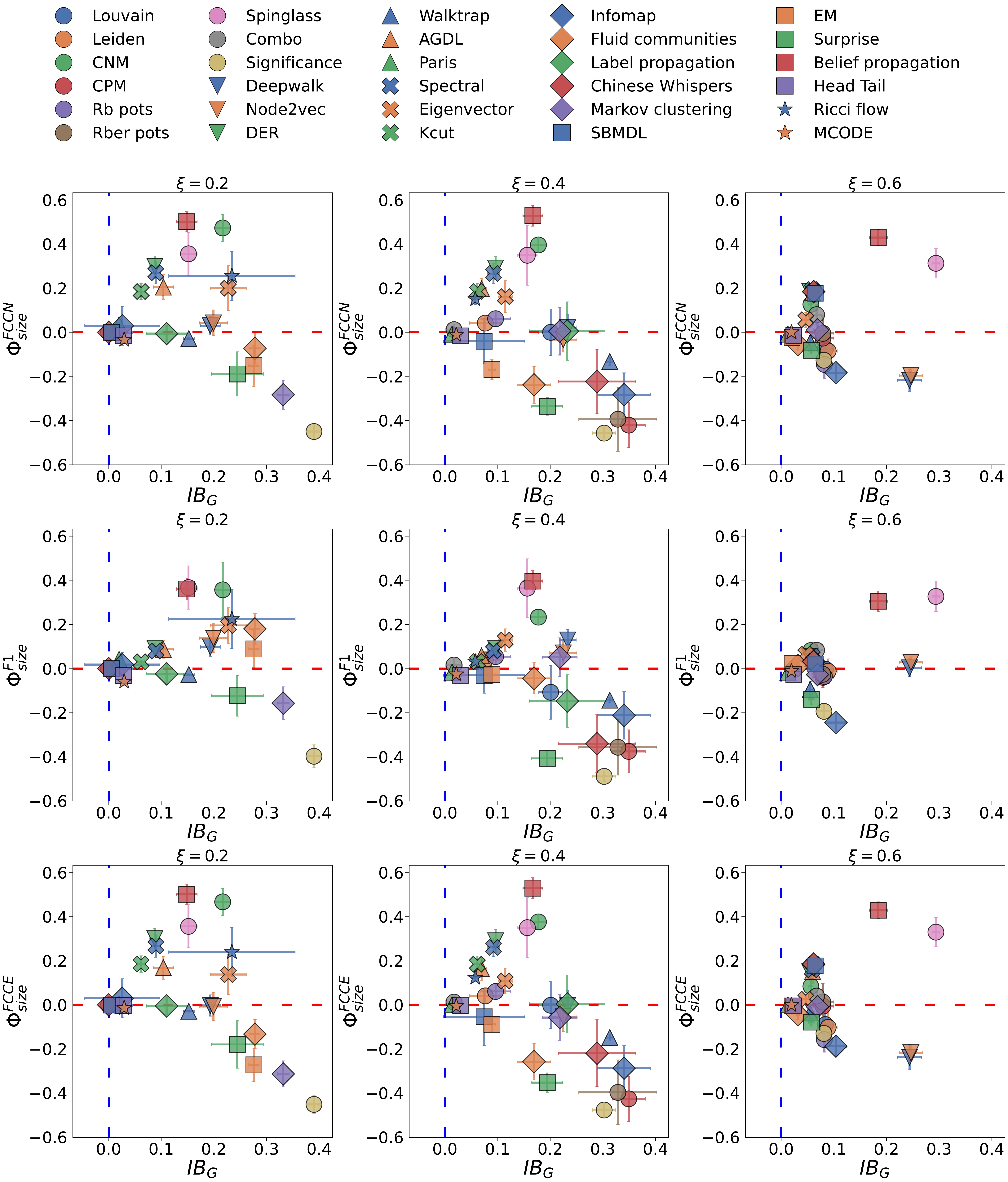}
    \caption{The relationship between the graph-level Individual Bias ($IB_G$) and the group fairness metric $\Phi_{\text{size}}$. The first row displays $IB_G$ versus $\Phi_{\text{size}}^{\text{FCCN}}$, the second row $IB_G$ versus $\Phi_{\text{size}}^{\text{F1}}$, and the third row $IB_G$ versus $\Phi_{\text{size}}^{\text{FCCE}}$. The red dashed line indicates perfect fairness according to $\Phi$, while the blue dashed line marks perfect fairness according to $IB_G$.}
    \label{fig:IVP_size}
\end{figure}

\begin{figure}[!t]
    \centering
    \includegraphics[width=\linewidth]{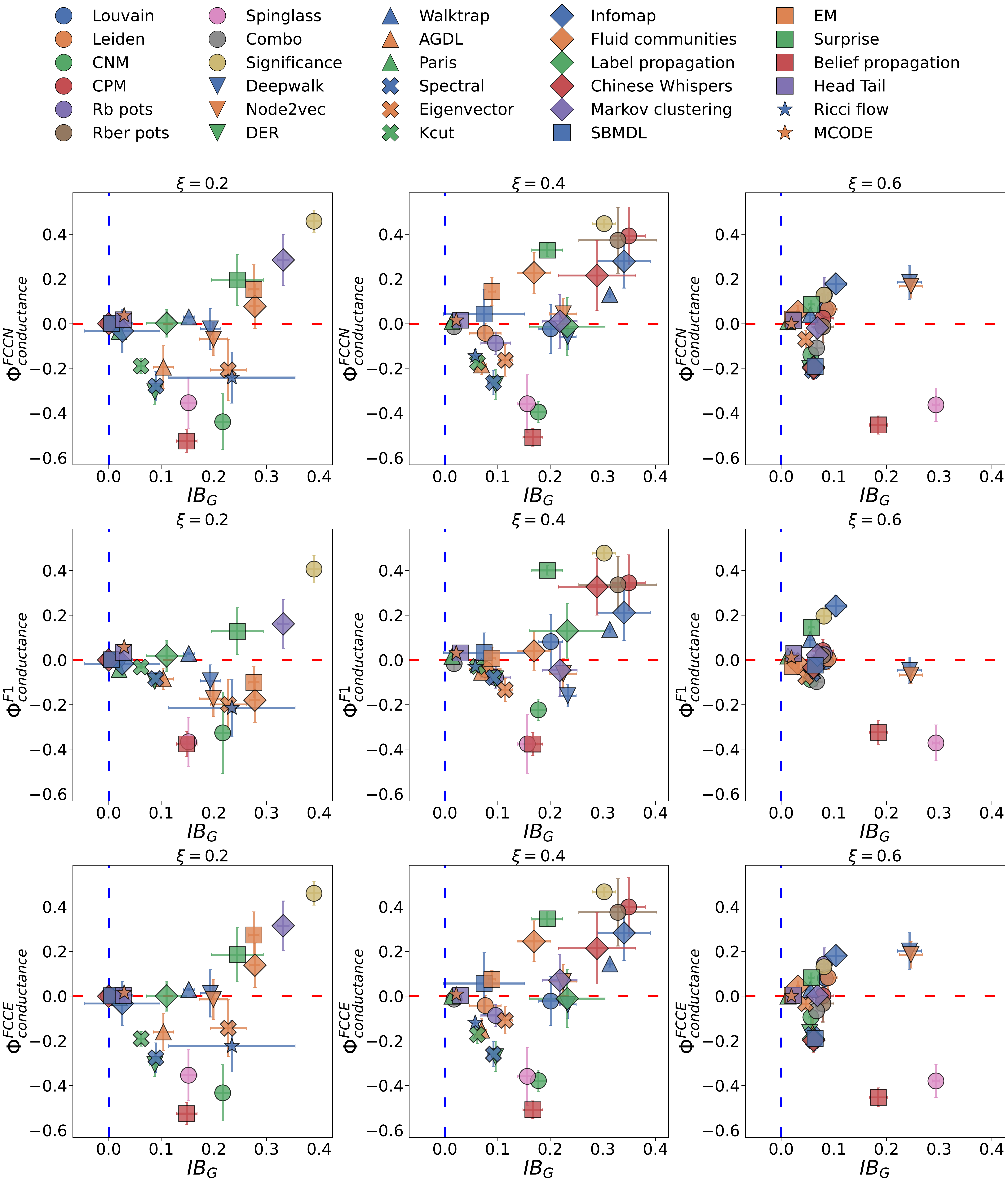}
    \caption{The relationship between the graph-level Individual Bias ($IB_G$) and the group fairness metric $\Phi_{\text{conductance}}$. The first row displays $IB_G$ versus $\Phi_{\text{conductance}}^{\text{FCCN}}$, the second row $IB_G$ versus $\Phi_{\text{conductance}}^{\text{F1}}$, and the third row $IB_G$ versus $\Phi_{\text{conductance}}^{\text{FCCE}}$. The red dashed line indicates perfect fairness according to $\Phi$, while the blue dashed line marks perfect fairness according to $IB_G$.}
    \label{fig:IVP_conductance}
\end{figure}

\begin{figure}[!t]
    \centering
    \includegraphics[width=\linewidth]{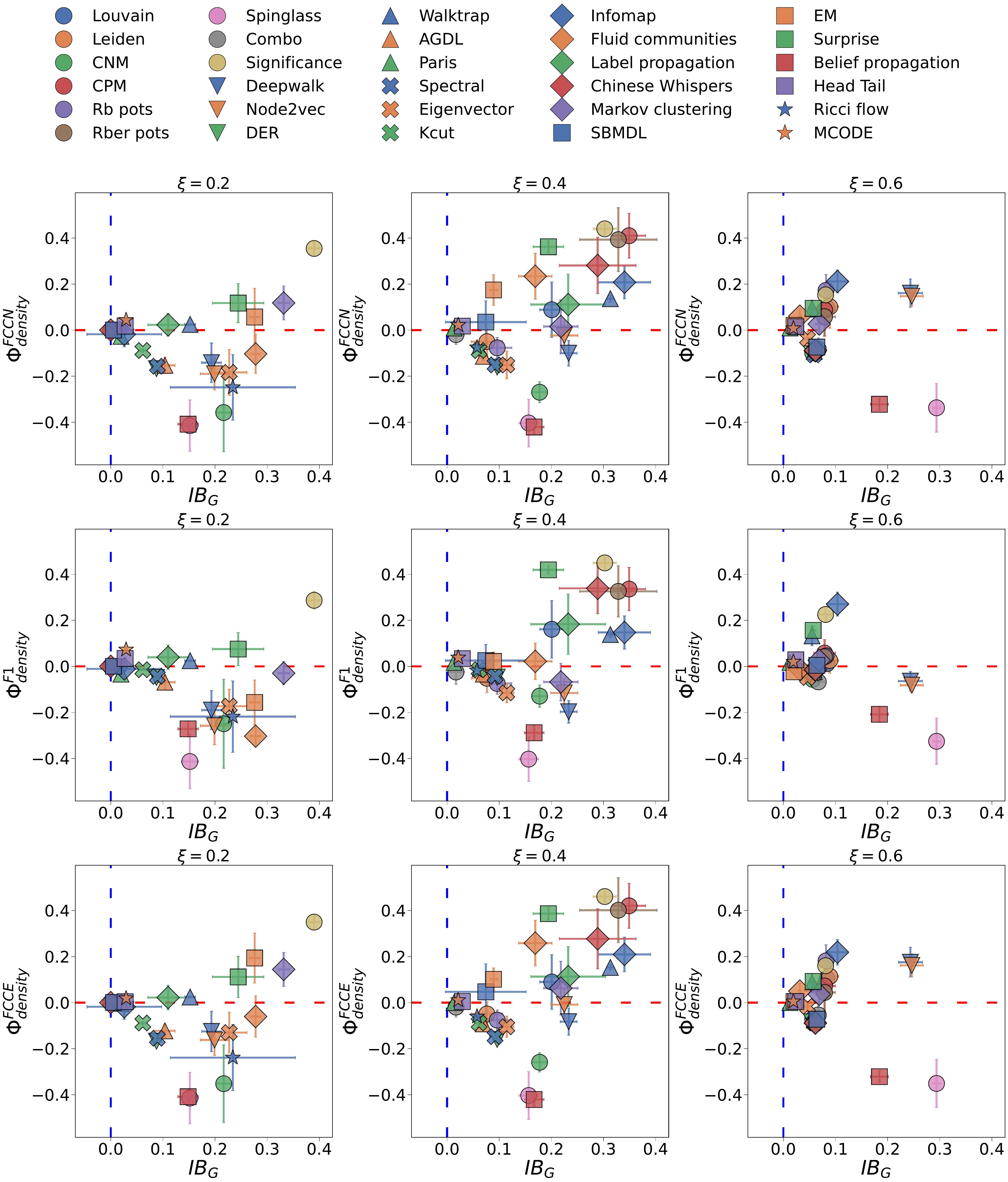}
    \caption{The relationship between the graph-level Individual Bias ($IB_G$) and the group fairness metric $\Phi_{\text{density}}$. The first row displays $IB_G$ versus $\Phi_{\text{density}}^{\text{FCCN}}$, the second row $IB_G$ versus $\Phi_{\text{density}}^{\text{F1}}$, and the third row $IB_G$ versus $\Phi_{\text{density}}^{\text{FCCE}}$. The red dashed line indicates perfect fairness according to $\Phi$, while the blue dashed line marks perfect fairness according to $IB_G$.}
    \label{fig:IVP_density}
\end{figure}

We compared our results for $IB_G$ with the group fairness metrics $\Phi$, that we described in Section~\ref{subsec:group_fairness_measure}.
In Figures~\ref{fig:IVP_size},~\ref{fig:IVP_conductance} and~\ref{fig:IVP_density}, we plot the $\Phi$ fairness measures (referring to the size, conductance, and density of communities, respectively) versus Individual fairness. In these plots, the dashed lines highlight the perfect fairness scores: red line for $IB_G$ and blue line for $\Phi$.
The methods that exhibit fair behaviour for both individual and group fairness measures are located near the intersection point of the two dotted lines.

In Figure~\ref{fig:IVP_size}, according to $\Phi_{size}$, the most unfair algorithms, especially in plots for $\xi=0.2$ and $\xi=0.4$, are on the lower side of the plots, meaning that the algorithms favour smaller size communities. The ones on the top side of the plot are the ones that favour larger communities. The algorithms that behave fairly for both individual and group fairness, for $\xi=0.2$, are SBMDL, Chinese Whispers, Infomap, Paris, Louvain, MCODE, and Head Tail: these are the same algorithms that were performing either perfectly or terribly in Section~\ref{subsec:ibg_results}. One surprising observation is that Significance is the most unfair algorithm for both categories, while de Vink et al.~\cite{devink2024groupfairnessmetricscommunity, de2025measuring} found out from their experiments that Significance tends to be a very fair method. A closer comparison reveals that this discrepancy stems from differences in the underlying network characteristics. Most datasets used in their study consist of dense graphs with many small communities, while our experiments employ sparse graphs with fewer but larger communities. The only comparable case with their analysis is the HICH-BA network, which shares a similar density and community structure with our benchmarks. As expected, it was the only instance where Significance exhibited unfair behaviour. After in-depth exploration, we can reasonably conclude that the Significance algorithm performs best on networks characterized by small, dense communities, but tends to behave unfairly in sparser, large-community settings. For $\xi=0.4$, the algorithms that are fair for both categories are Paris, Head Tail, MCODE, and Combo, similarly to the results for $\xi=0.2$. The other fair algorithms for both $IB_G$ and $\Phi_{size}$ for $\xi=0.2$ tend to remain somewhat fair group-wise, while they become unfair individually speaking. For $\xi=0.6$, we notice a collapse towards the intersection between the two fair lines. We can notice that a few algorithms behave extremely unfairly: Spinglass, Belief propagation, Deepwalk, and Node2Vec are unfair for both individual and group fairness. Deepwalk and Node2Vec are estimated as fair for $\Phi_{size}^{F1}$.

In Figure~\ref{fig:IVP_conductance} and in Figure~\ref{fig:IVP_density}, the plots show that the most unfair algorithms are on the top side of the plots, differently from what we saw in Figure~\ref{fig:IVP_size}, and it becomes more noticeable in the plots for $\xi=0.4$. This means that, the CD algorithms for our data favour higher conductance and higher density communities. Looking closer, most of the algorithms that were shown to have scored a negative $\Phi_{size}$ show positive $\Phi_{conductance}$ and positive $\Phi_{density}$. Also, the same stands for positive $\Phi_{size}$, that translates to negative $\Phi_{conductance}$ and negative $\Phi_{density}$.

In Figure~\ref{fig:IVP_conductance}, we can find the same cluster from Figure~\ref{fig:IVP_size}, speaking of algorithms that are fair for both categories: SBMDL, Chinese Whispers, Infomap, Paris, Louvain, MCODE, and Head Tail. Significance is still the most unfair algorithm for $\xi=0.2$, favouring higher conductance communities. $\Phi_{conductance}^{F1}$ tends to underestimate the group unfairness for most methods, compared to $\Phi_{conductance}^{FCCN}$ and $\Phi_{conductance}^{FCCE}$, except for Fluid Communities. In fact, in the plots for $FCCN$ and $FCCE$, it is close to $0.1$, but in the plot for $F1$, it is close to $0.3$. The Same thing happens in plots for $\xi=0.4$, even though Fluid Communities' unfairness switches from favouring low conductance communities to favouring high conductance communities.

The results shown in Figures~\ref{fig:IVP_conductance} and~\ref{fig:IVP_density} are very similar, except for some minor differences, like Deepwalk and Node2vec being fairer for $\Phi_{conductance}$ for both $\xi=0.2$ and $\xi=0.4$. The differences become even less significant if we consider the plots for $\xi=0.6$. Spinglass unfairness, according to $\Phi_{density}$ is overestimated with respect to the values in Figure~\ref{fig:IVP_size} and~\ref{fig:IVP_conductance}. For $\xi=0.6$, the observations are very similar to the ones regarding Figure~\ref{fig:IVP_size}, with the only difference that the algorithms on the positive part of the plot are on the negative part, and vice versa.

In Figure~\ref{fig:IVP_conductance} and~\ref{fig:IVP_size} we can pinpoint some algorithms that generally show a good group-level fairness while being individually unfair, but the most noticeable is Label Propagation. The only plots where it is further away from the red dotted line are in the $\Phi^{F1}$ plots, where it shows that it favours smaller communities with high conductance.

Generally, we observe that all the methods considered group-wise unfair are also considered unfair from an individual perspective. As already noticed in~\ref{tab:IBg_results} for $IB_G$, also for $\Phi$, the number of unfair algorithms diminishes as the $\xi$ value increases. Notably, there are no algorithms lying directly on, or close to, the blue dotted line, except at its intersection with the red dotted line. This indicates that an algorithm can be group-wise fair without being individually fair, but the opposite is not true. This is due to the measures' different behaviours in response to local perturbations: when a node is assigned to an incorrect community, it creates individual unfairness not only for itself, but also for every member of its ground-truth community, as well as the community it is wrongly assigned to. This perturbation is significant from an individual perspective, yet remains relatively minor at the group level. Even when an entire community is absorbed by other communities, the overall impact on $\Phi$ remains statistically irrelevant if we consider a sufficient number of communities. This observation aligns with the observations presented in Section~\ref{subsec:ibg_results}, where we noticed that as the algorithms' performances tend towards medium values, $IB_G$ gets to its highest values. Therefore, while a CD algorithm may achieve group fairness without ensuring individual fairness, the converse cannot hold, as no algorithm can be individually fair (according to $IB$) without also being fair at the group level.

To summarize, we can conclude that there is not a strict correlation between Individual and Group fairness, since the latter can be interpreted as an aggregation of the former, shifting the focus from single entities to groups. Consequently, many details concerning bias and the unfair treatment of individuals become diluted in favour of a broader, more community-oriented perspective.

\subsection{Individual Bias at Node-level: Distribution Analysis}\label{subsec:ib_distribution_analysis}

\begin{figure}[!htb]
    \centering
    \begin{subfigure}[b]{.32\linewidth}
        \centering
        \includegraphics[width=\linewidth]{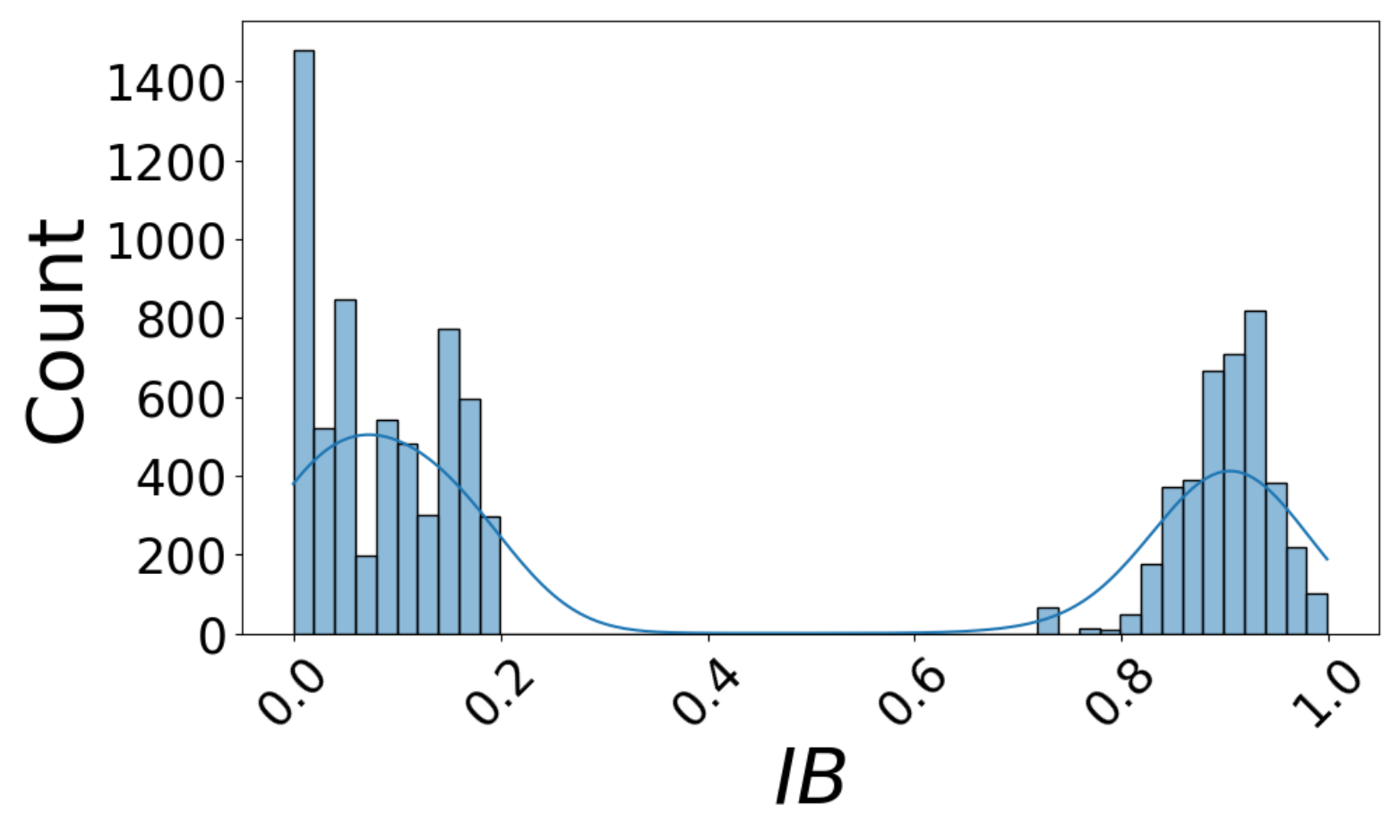}
        \caption{}
        \label{fig:IB_significance_02}
    \end{subfigure}
    \begin{subfigure}[b]{.32\linewidth}
        \centering
        \includegraphics[width=\linewidth]{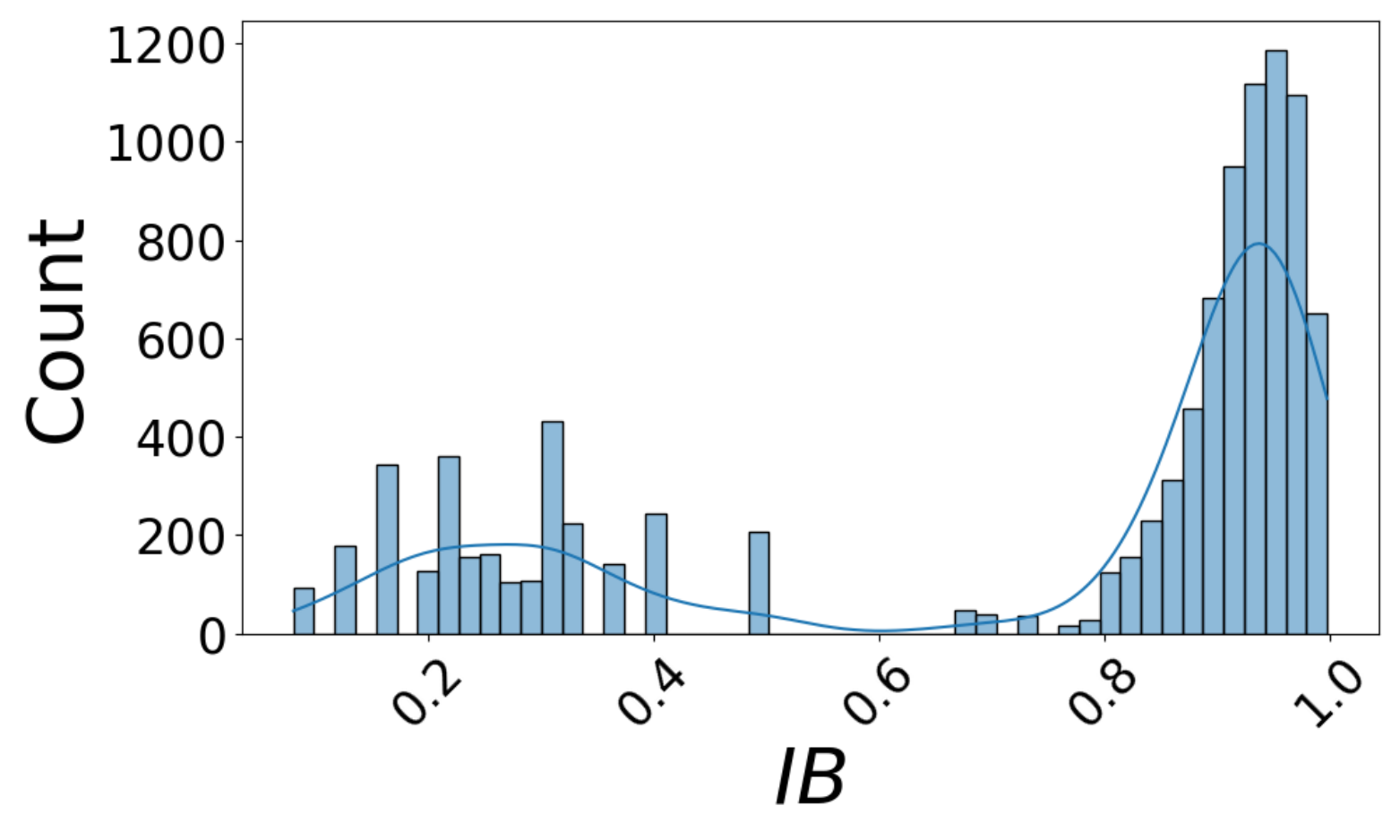}
        \caption{}
        \label{fig:IB_significance_04}
    \end{subfigure}
    \begin{subfigure}[b]{.32\linewidth}
        \centering
        \includegraphics[width=\linewidth]{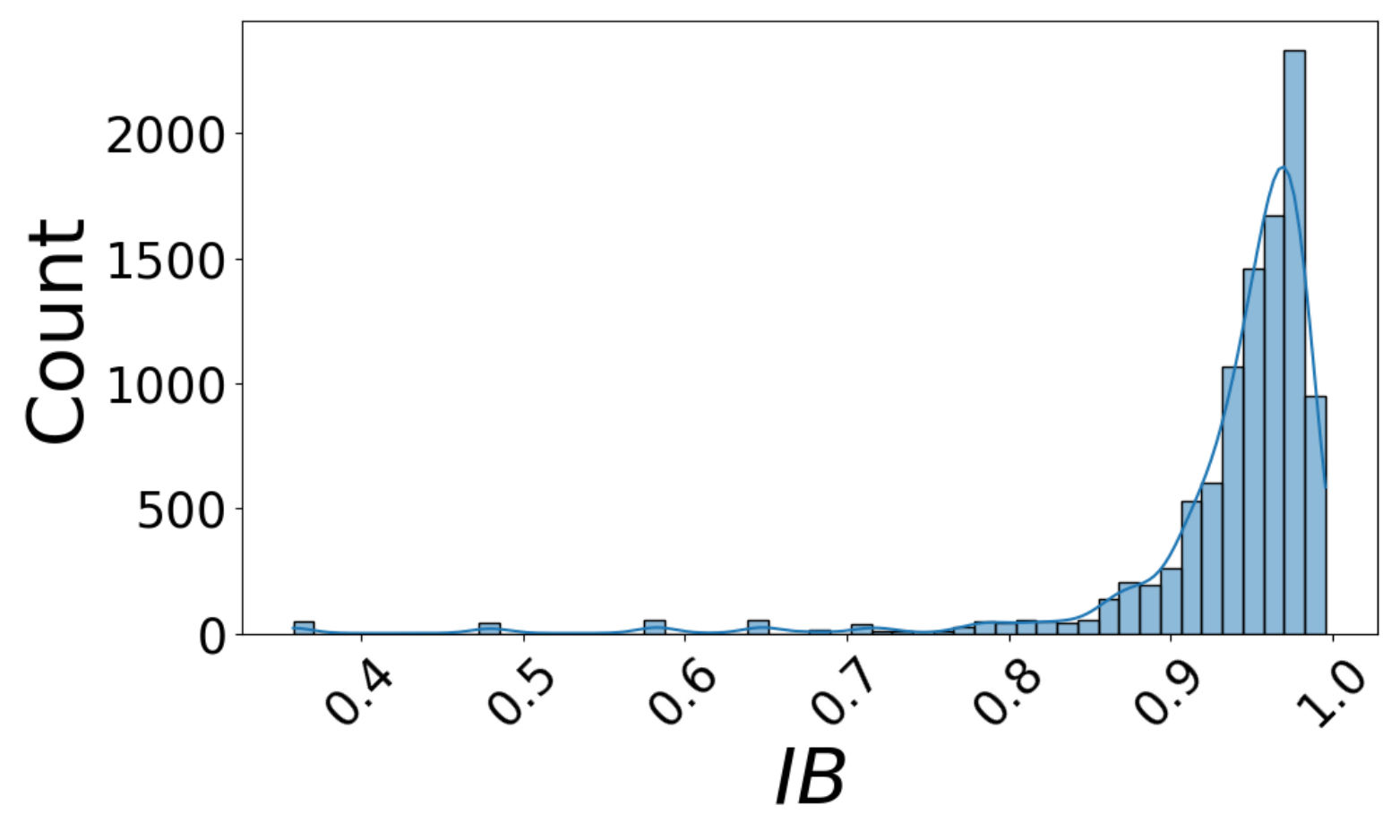}
        \caption{}
        \label{fig:IB_significance_06}
    \end{subfigure}
    \begin{subfigure}[b]{.32\linewidth}
        \centering
        \includegraphics[width=\linewidth]{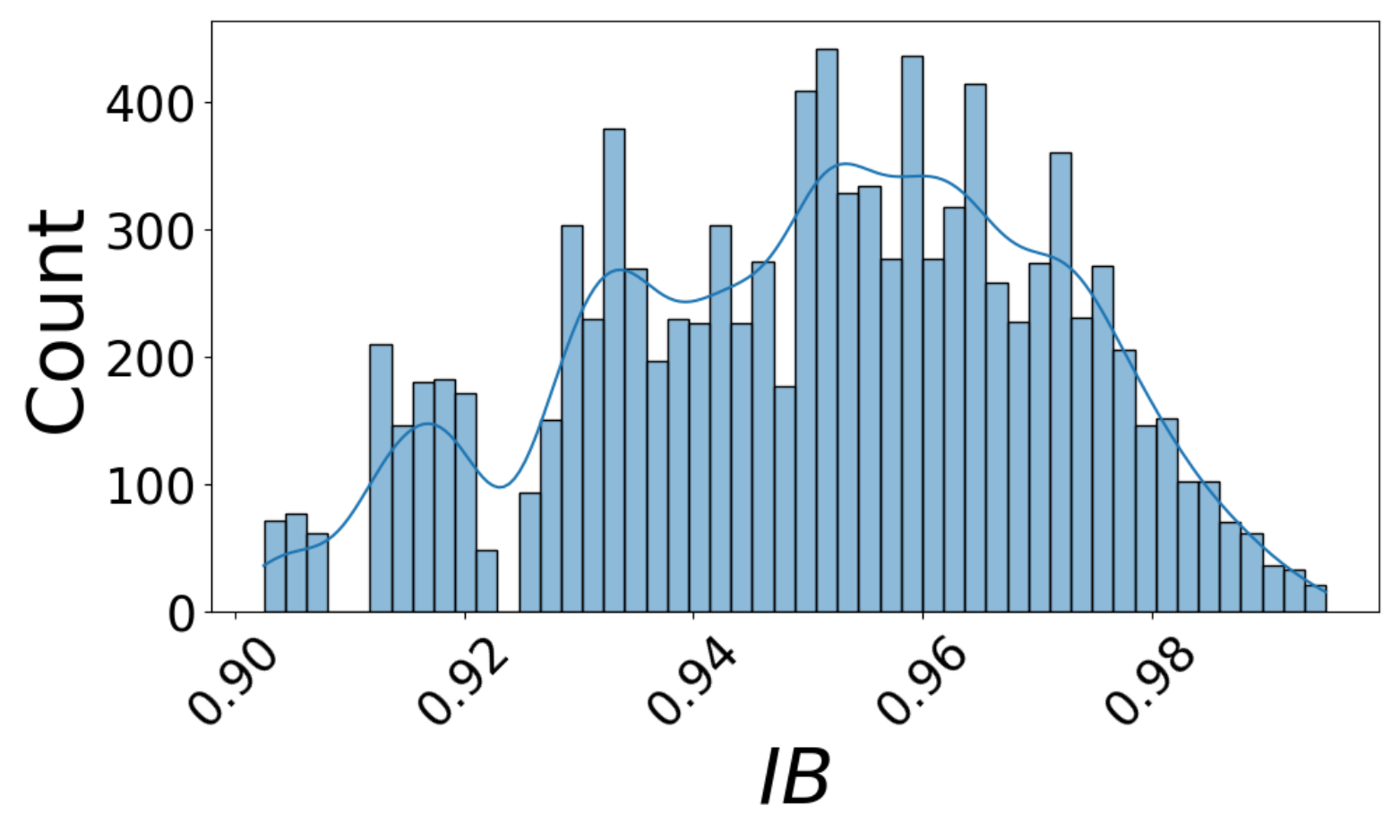}
        \caption{}
        \label{fig:IB_paris_02}
    \end{subfigure}
    \begin{subfigure}[b]{.32\linewidth}
        \centering
        \includegraphics[width=\linewidth]{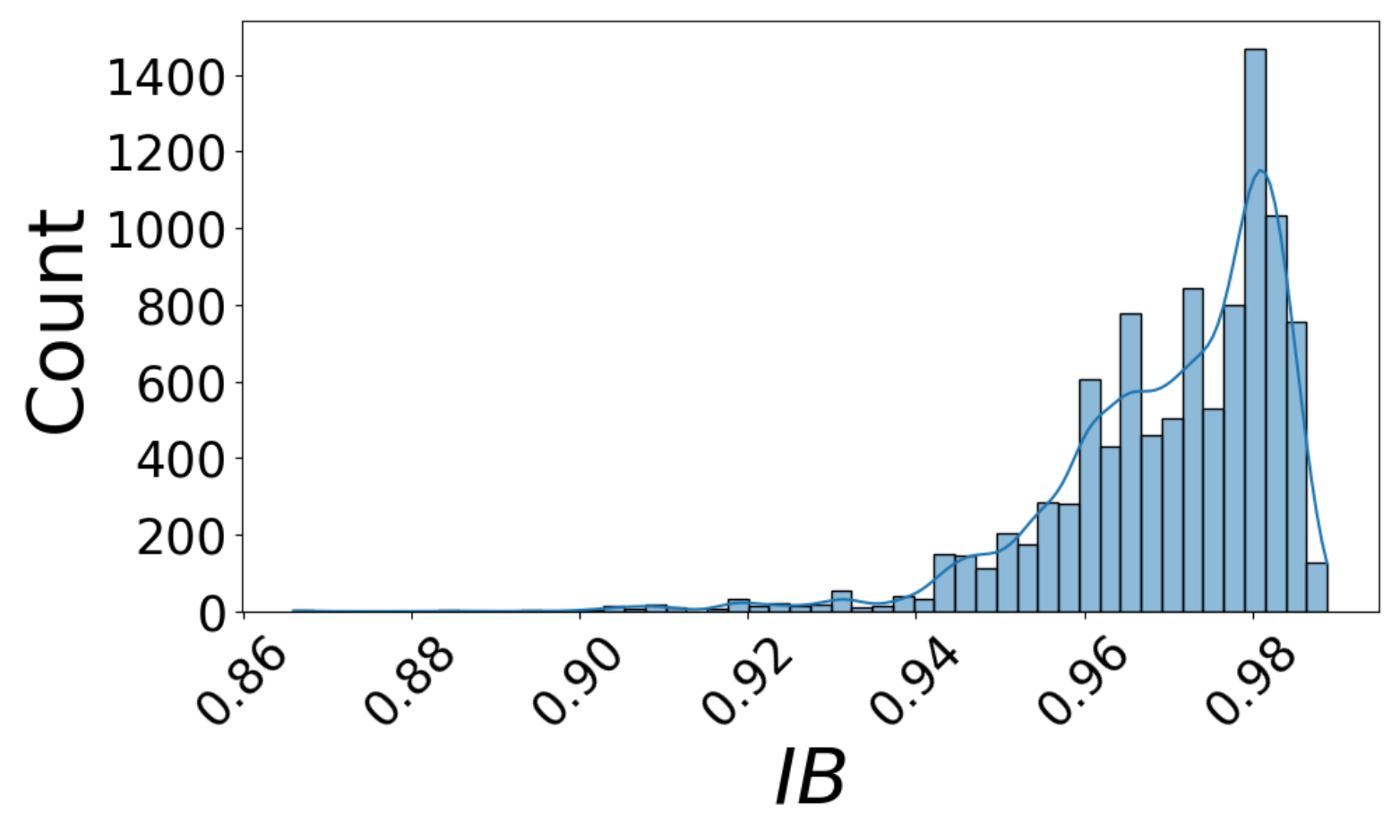}
        \caption{}
        \label{fig:IB_paris_04}
    \end{subfigure}
    \begin{subfigure}[b]{.32\linewidth}
        \centering
        \includegraphics[width=\linewidth]{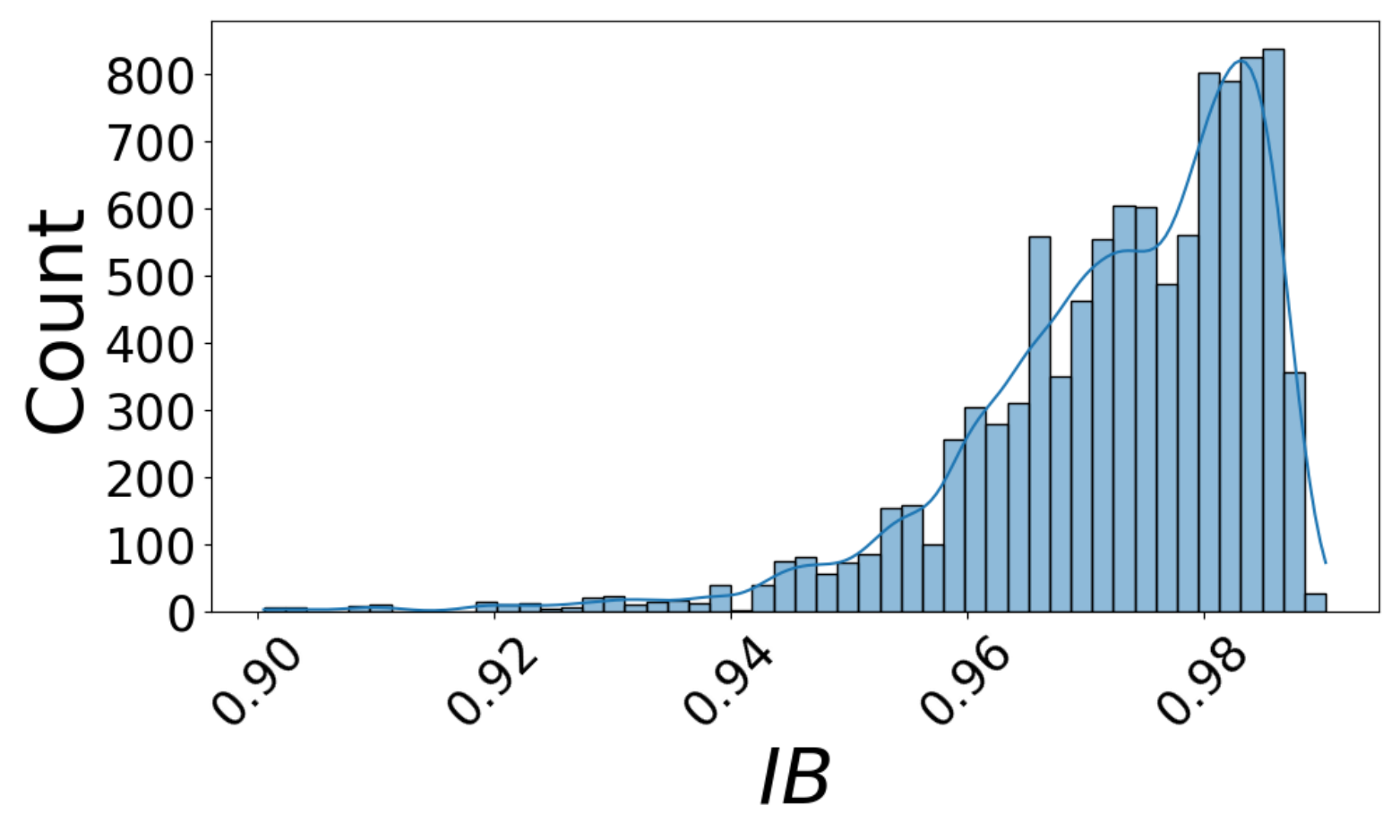}
        \caption{}
        \label{fig:IB_paris_06}
    \end{subfigure}
    \begin{subfigure}[b]{.32\linewidth}
        \centering
        \includegraphics[width=\linewidth]{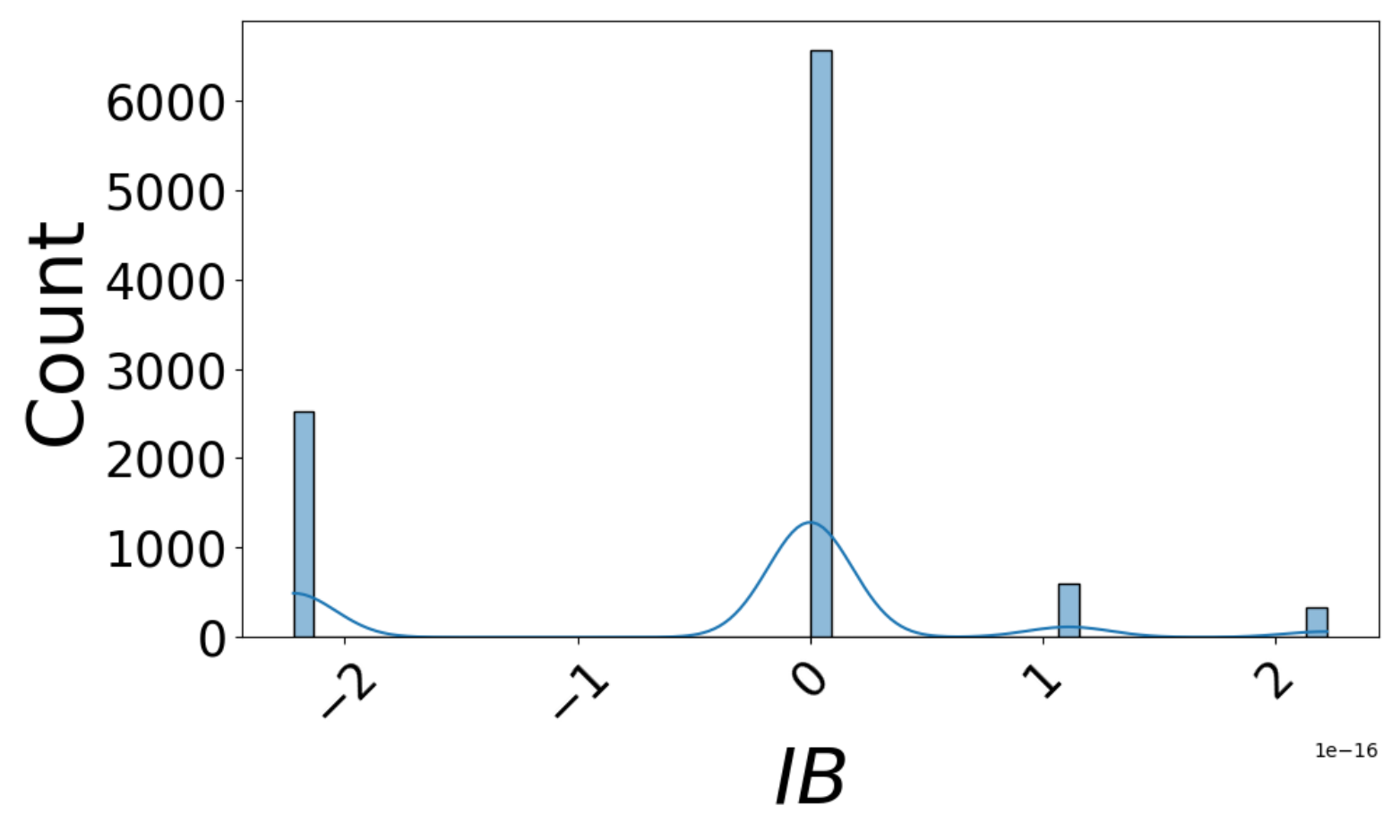}
        \caption{}
        \label{fig:IB_chinese_02}
    \end{subfigure}
    \begin{subfigure}[b]{.32\linewidth}
        \centering
        \includegraphics[width=\linewidth]{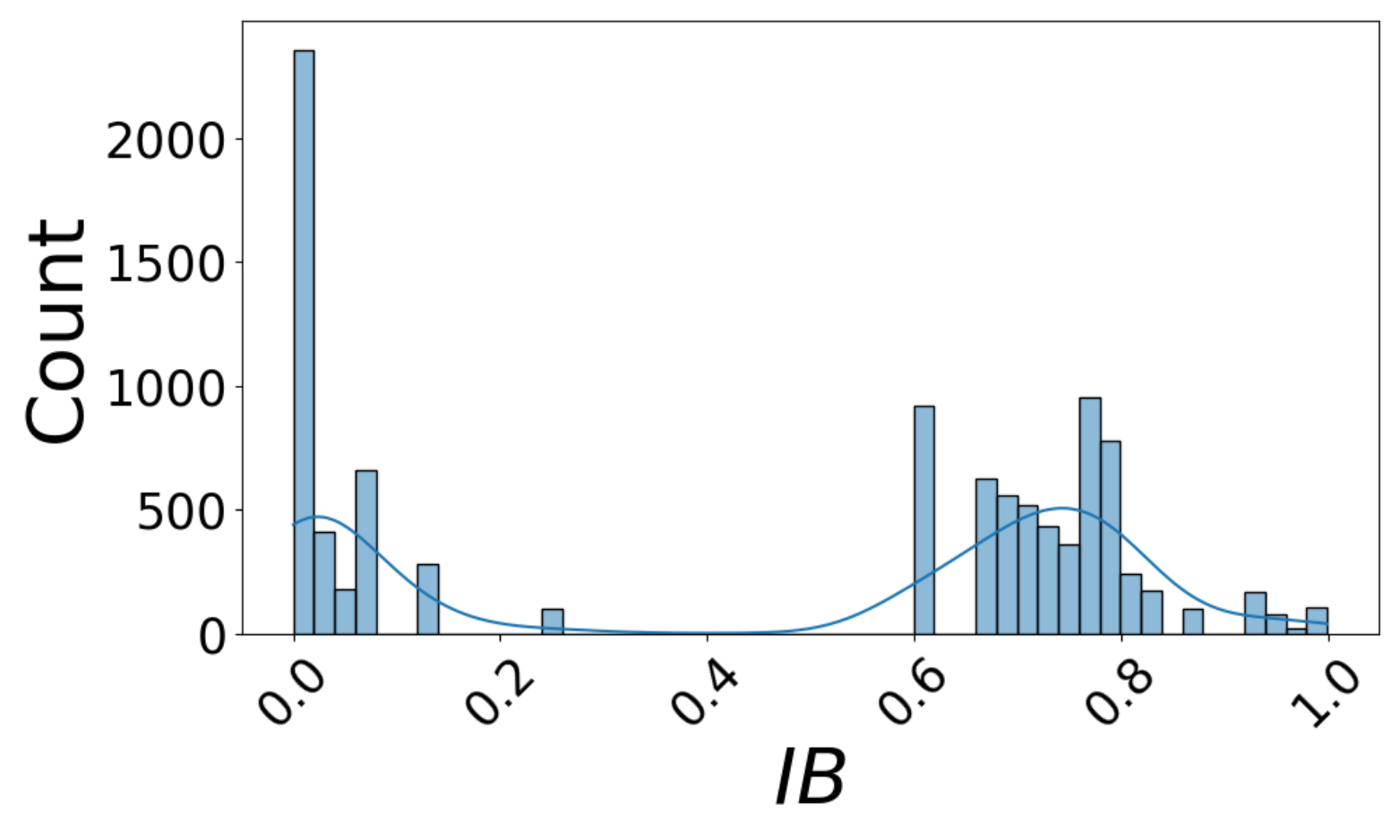}
        \caption{}
        \label{fig:IB_chinese_04}
    \end{subfigure}
    \begin{subfigure}[b]{.32\linewidth}
        \centering
        \includegraphics[width=\linewidth]{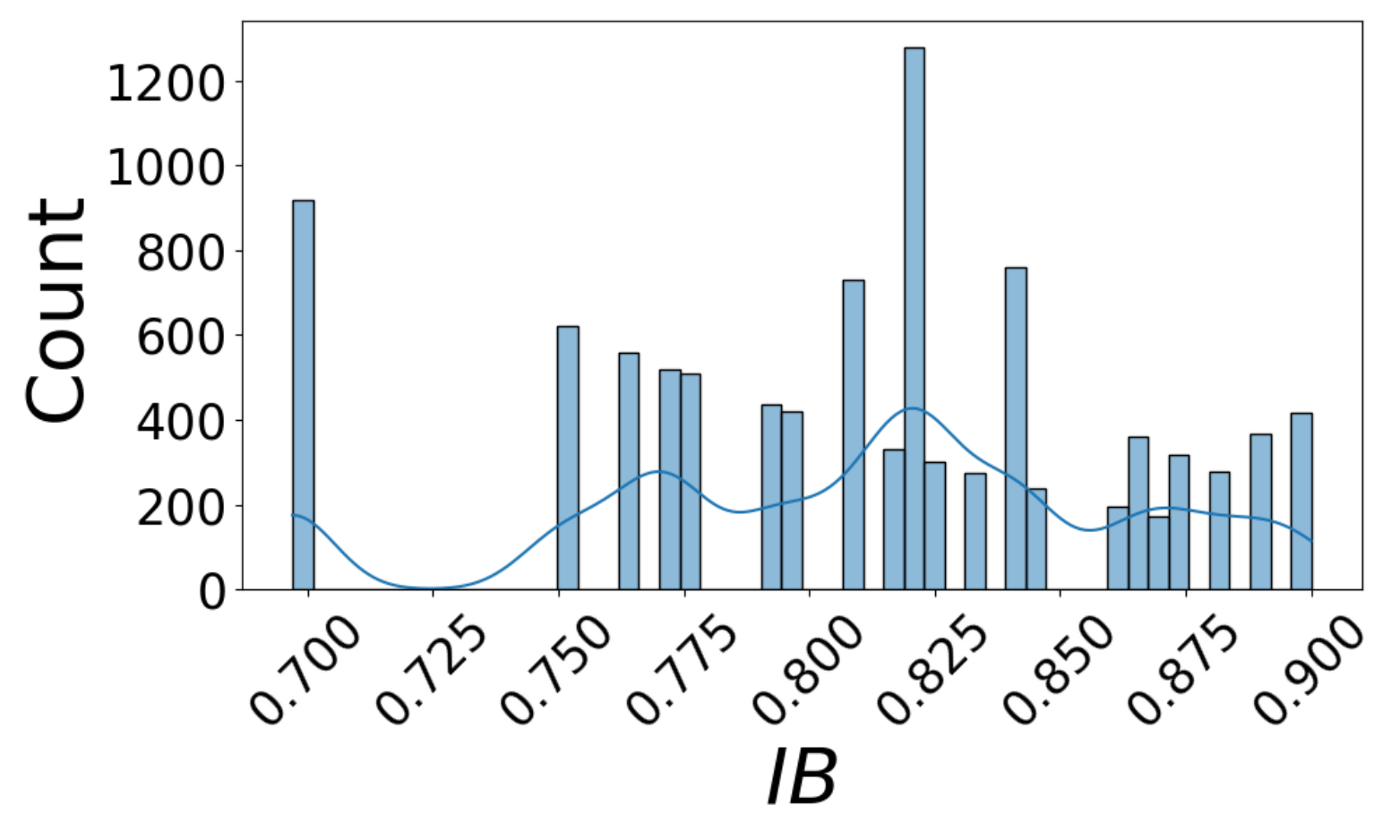}
        \caption{}
        \label{fig:IB_chinese_06}
    \end{subfigure}
    \begin{subfigure}[b]{.32\linewidth}
        \centering
        \includegraphics[width=\linewidth]{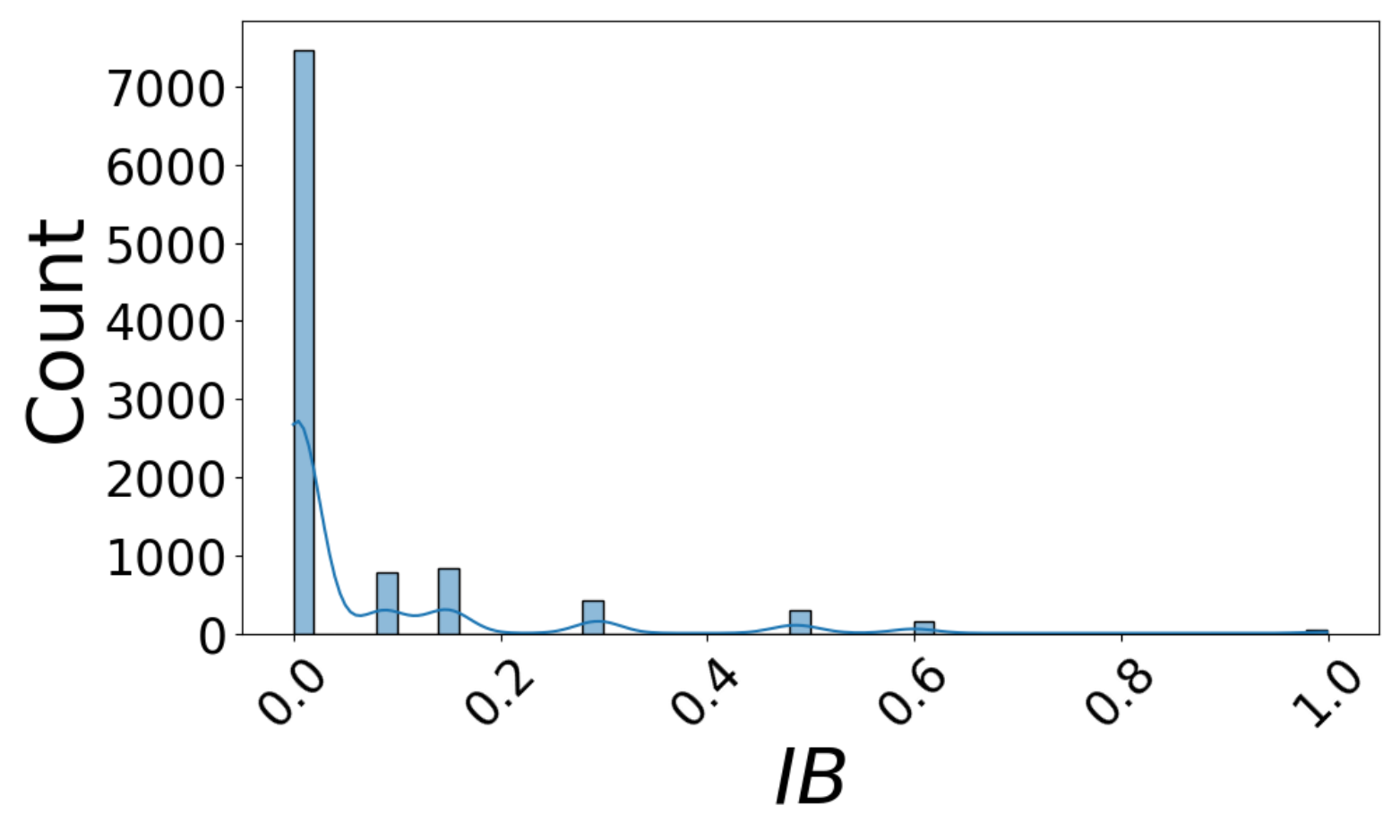}
        \caption{}
        \label{fig:IB_spinglass_02}
    \end{subfigure}
    \begin{subfigure}[b]{.32\linewidth}
        \centering
        \includegraphics[width=\linewidth]{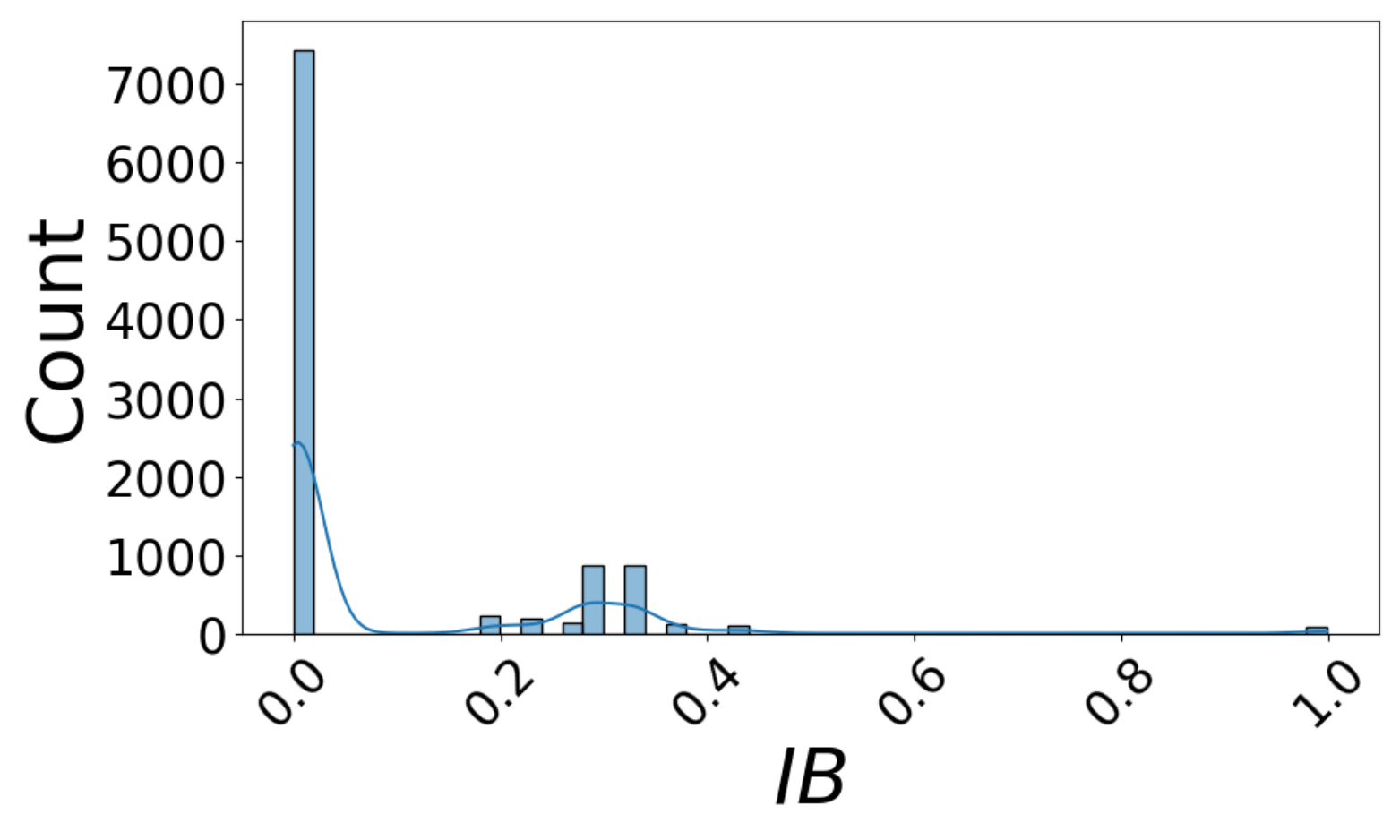}
        \caption{}
        \label{fig:IB_spinglass_04}
    \end{subfigure}
    \begin{subfigure}[b]{.32\linewidth}
        \centering
        \includegraphics[width=\linewidth]{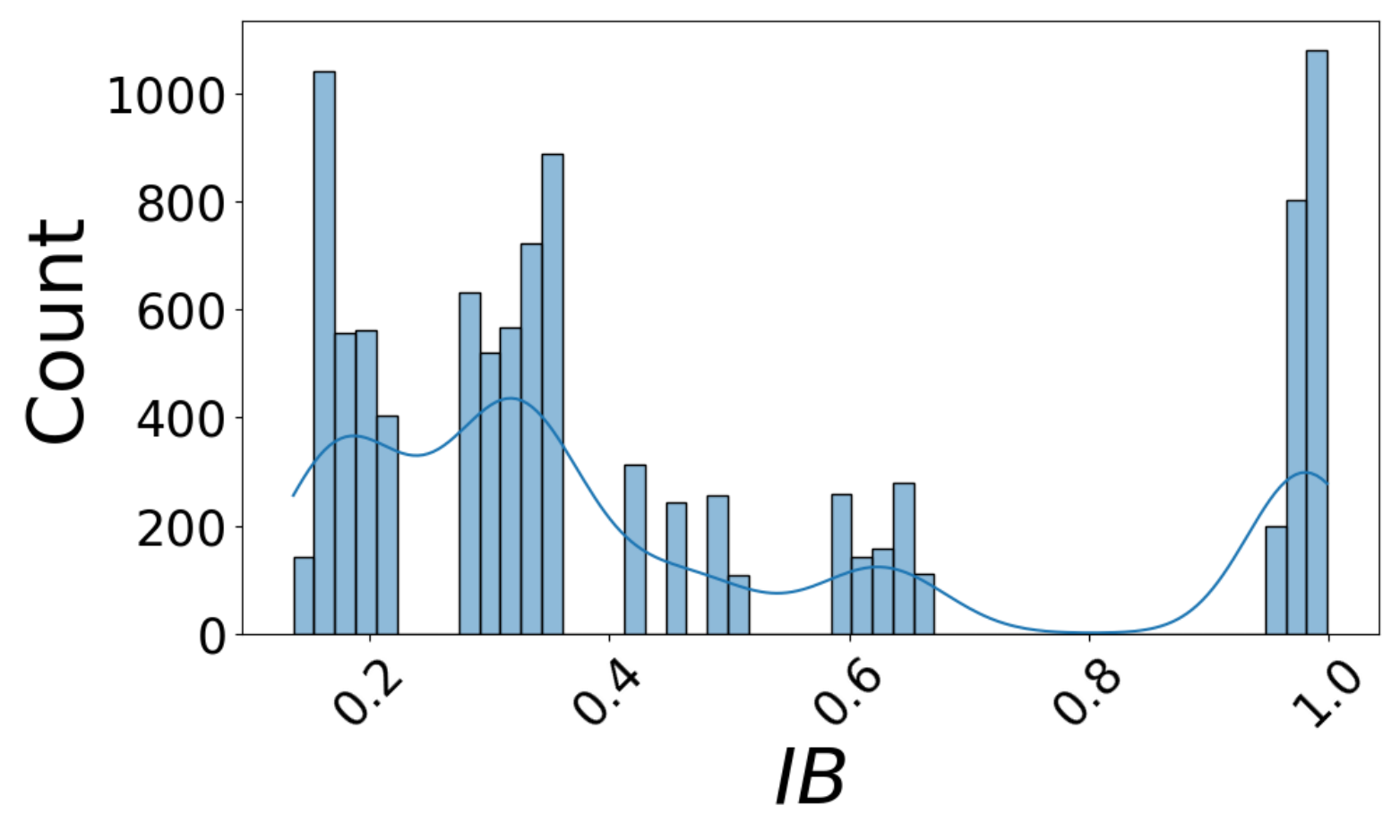}
        \caption{}
        \label{fig:IB_spinglass_06}
    \end{subfigure}
    \caption{Distributions of $IB$ values computed on different algorithms' communities' clustering results for ABCD network. The algorithms' results shown are: Significance ($\xi=0.2$ (a), $\xi=0.4$ (b), and $\xi=0.6$ (c)), Paris ($\xi=0.2$ (d), $\xi=0.4$ (e), and $\xi=0.6$ (f)), Chinese Whispers ($\xi=0.2$ (g), $\xi=0.4$ (h), and $\xi=0.6$ (i)), and Spinglass ($\xi=0.2$ (j), $\xi=0.4$ (k), and $\xi=0.6$ (l)).}
    \label{fig:IB_distributions}
\end{figure}

Here, we analyse the distribution of node-level $IB$ values for a subset of algorithms selected to illustrate distinct fairness behaviours: Significance Communities, Paris, Chinese Whispers, and Spinglass (as shown in Figure~\ref{fig:IB_distributions}) across all values of $\xi$.

Paris consistently showed low fairness and poor clustering quality in earlier results, and the $IB$ distributions clarify why. Across all $\xi$ values, almost all nodes have $IB > 0.9$ (with a few nodes above 0.85 at $\xi=0.4$). Thus, Paris treats every node uniformly poorly, leading to high individual unfairness for each node but low graph-level unfairness ($IB_G$), since all nodes experience the same degree of context change. The reason for Paris' poor performance is to be searched in the sparseness of our dataset and the structure of ABCD graphs. The presence of big hubs and relatively sparse communities can lead to early mistakes in the agglomerative process of the Paris algorithm, causing a cascade effect that will impact the overall CD process.

Significance method has high individual bias as shown in (Figure~\ref{fig:IB_significance_02},~\ref{fig:IB_significance_04}, and~\ref{fig:IB_significance_06}). At $\xi=0.2$, the $IB$ distribution is highly polarised: roughly half the nodes are treated very unfairly, indicating misclassification or substantial contextual distortion, while the others are treated almost perfectly fairly. As $\xi$ increases, the distribution shifts rightwards, with unfairness spreading across a larger portion of the graph. At $\xi=0.6$, most nodes are unfairly classified, producing a left-skewed distribution. This aligns with our earlier findings (Section~\ref{subsec:ibg_results}) that Significance tends to overestimate the number of communities in sparse networks. For this reason, the local context of many nodes has drastically changed.

Chinese Whispers exhibits irregular $IB_G$ values across $\xi$, and the distributions reveal the cause. At $\xi=0.2$, all nodes have $IB=0$, reflecting perfect fairness. At $\xi=0.4$, the distribution becomes bimodal, with many nodes remaining fair but a substantial group receiving $IB \geq 0.6$. At $\xi=0.6$, all nodes have high individual bias ($0.7 \leq IB \leq 0.9$), producing a low $IB_G$ despite uniformly poor treatment. This happens because, as communities become harder to detect, Chinese whispers tends to add all nodes to one big community and divide the remaining nodes into smaller ones. So we go from a nearly perfect prediction for $\xi=0.2$, to a big cluster with a few small communities for $\xi=0.4$, to all nodes in the same community for $\xi=0.6$. The $IB_G$ values, therefore, reflect the shift between perfect fairness and uniformly unfair behaviour.

Spinglass presents the clearest fairness–performance trade-off, especially for ABCD graphs with $\xi=0.6$ (Figure~\ref{fig:IB_spinglass_06}). While many nodes receive very low $IB$, indicating correct or near-correct community assignments, a non-trivial fraction of nodes are treated unfairly. For $\xi=0.6$, the algorithm still retains a good performance because many nodes are assigned to their correct community, but a high number of nodes are treated very unfairly, with more than 10\% of them having $IB\approx1.0$. This mixture of fair and highly unfair treatment explains the high $IB_G$ values and Spinglass’ unique placement in earlier scatter plots.

Overall, the distribution patterns confirm that individually unfair algorithms show polarised or heavy-tailed $IB$ distributions, while individually fair algorithms exhibit tightly concentrated distributions, even when fairness manifests through uniformly poor treatment.

\begin{figure}[!t]
    \centering
    \begin{subfigure}[b]{\linewidth}
        \centering
        \includegraphics[width=.83\linewidth]{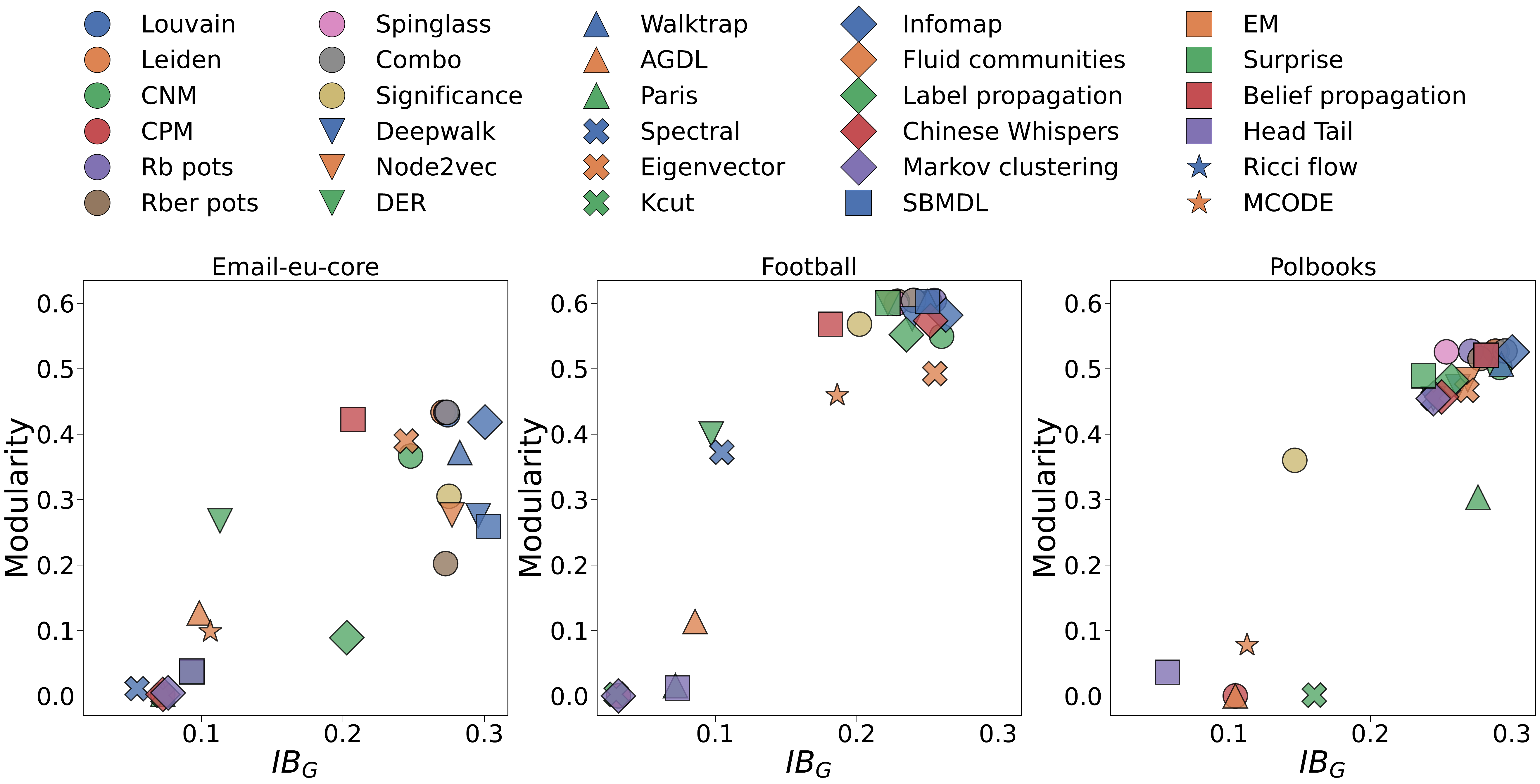}
        \label{fig:RW_IBg_vs_Modularity}
    \end{subfigure}
    \begin{subfigure}[b]{\linewidth}
        \centering
        \includegraphics[trim={0 0 0 14cm}, clip, width=.83\linewidth]{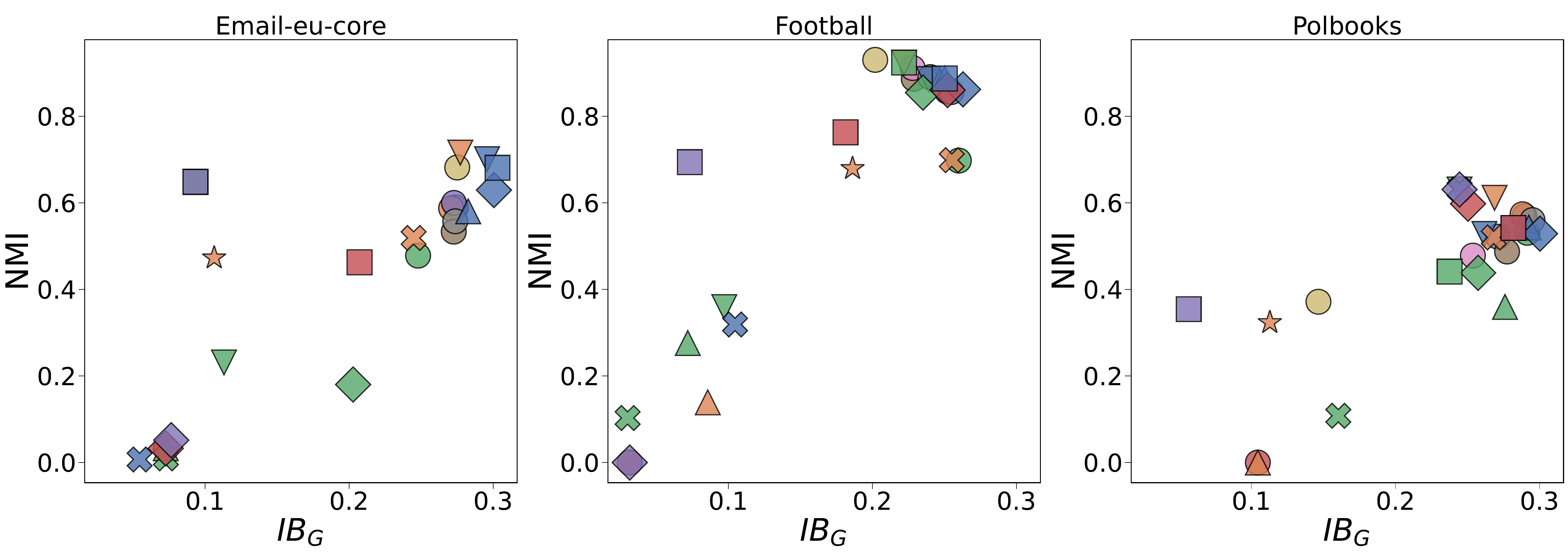}
        \label{fig:RW_IBg_vs_NMI}
    \end{subfigure}
    \begin{subfigure}[b]{\linewidth}
        \centering
        \includegraphics[trim={0 0 0 14cm}, clip, width=.83\linewidth]{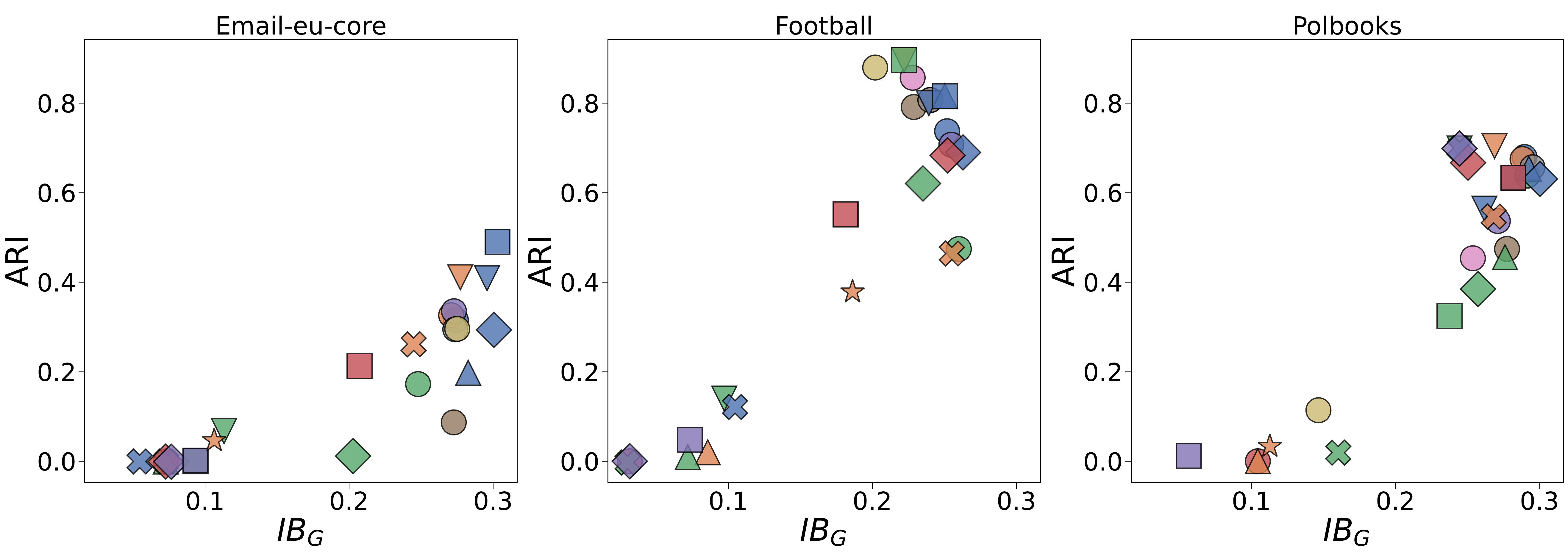}
        \label{fig:RW_IBg_vs_ARI}
    \end{subfigure}
    \begin{subfigure}[b]{\linewidth}
        \centering
        \includegraphics[trim={0 0 0 14cm}, clip, width=.83\linewidth]{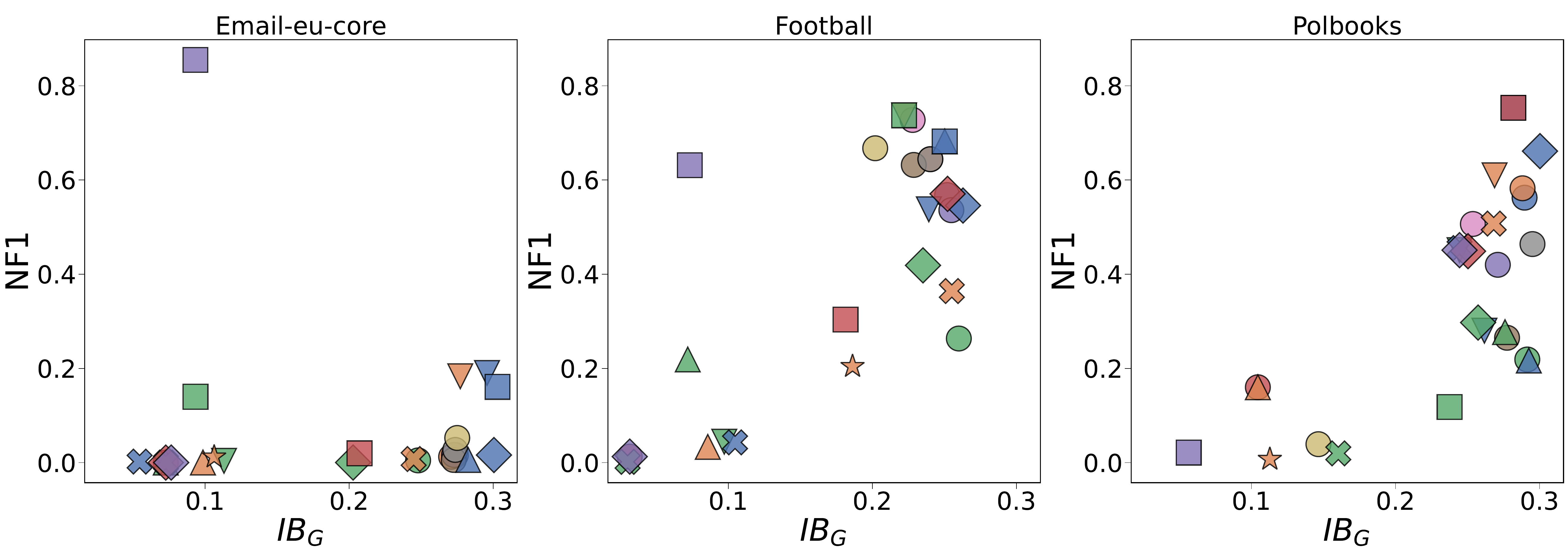}
        \label{fig:RW_IBg_vs_NF1}
    \end{subfigure}
    \caption{$IB_G$ measures against communities' quality scores for real-world networks.}
    \label{fig:RW_IBg_vs_performance}
\end{figure}

\subsection{Results on Real-world Networks}\label{subsec:real_world_res}

\begin{figure}[!htb]
    \centering
    \includegraphics[width=\linewidth]{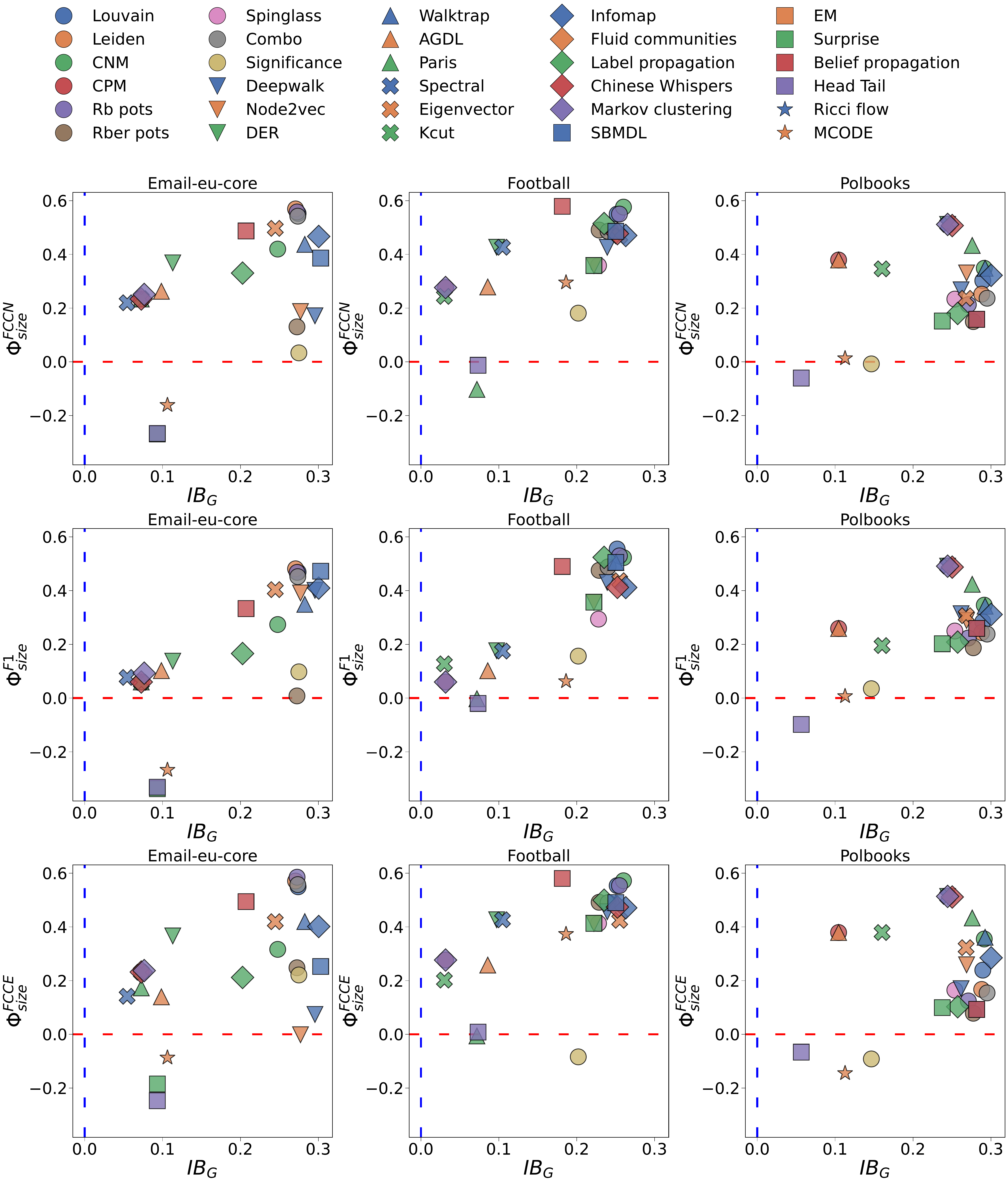}
    \caption{$IB_G$ measure against $\Phi_{size}$ measures for real-world networks. The red dotted line is the perfect fairness for $\Phi$, the blue dotted line is the perfect fairness for $IB_G$.}
    \label{fig:RW_IVP_size}
\end{figure}

\begin{figure}[!htb]
    \centering
    \includegraphics[width=\linewidth]{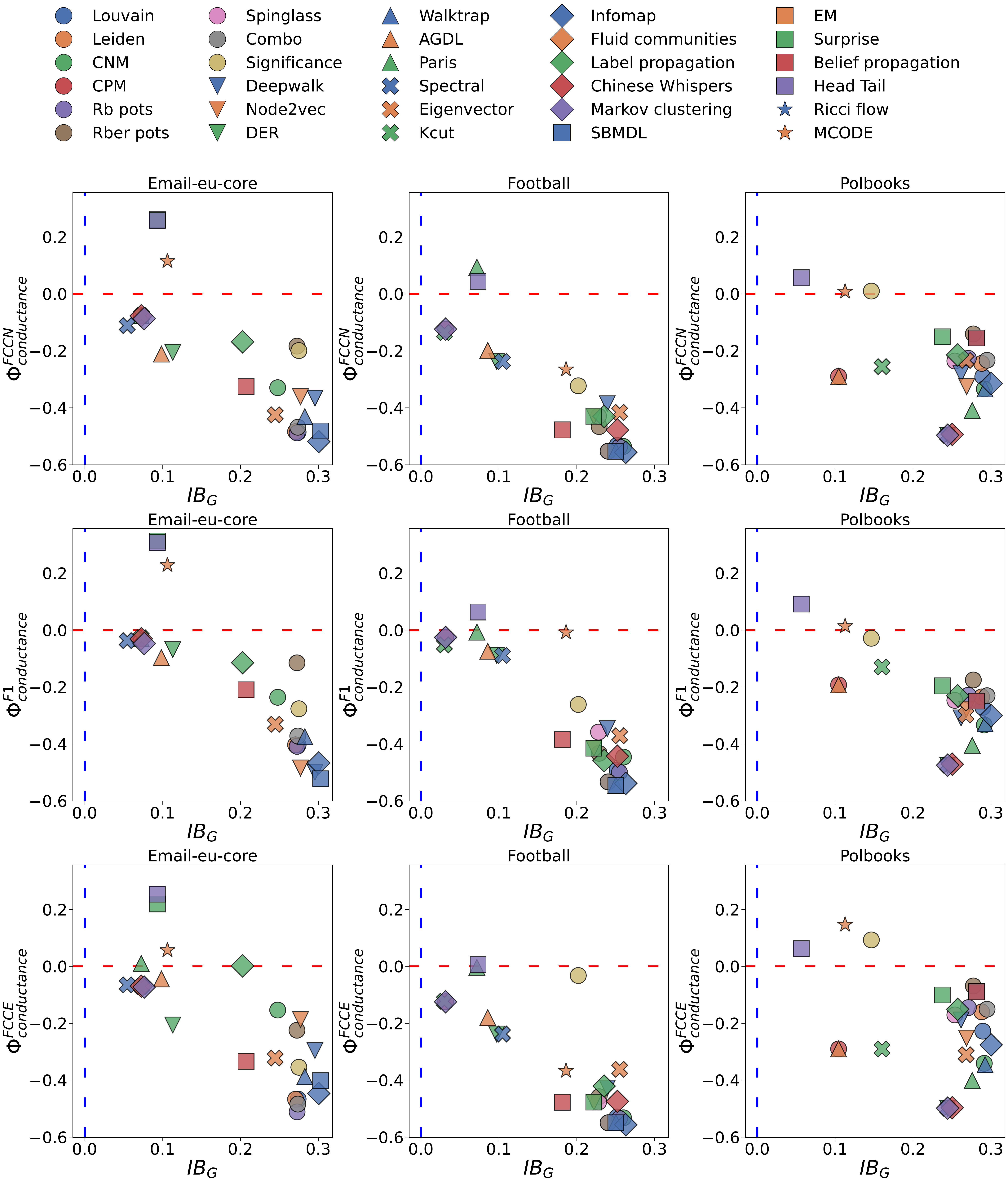}
    \caption{$IB_G$ measure against $\Phi_{conductance}$ measures for real-world networks. The red dotted line is the perfect fairness for $\Phi$, the blue dotted line is the perfect fairness for $IB_G$.}
    \label{fig:RW_IVP_conductance}
\end{figure}

\begin{figure}[!htb]
    \centering
    \includegraphics[width=\linewidth]{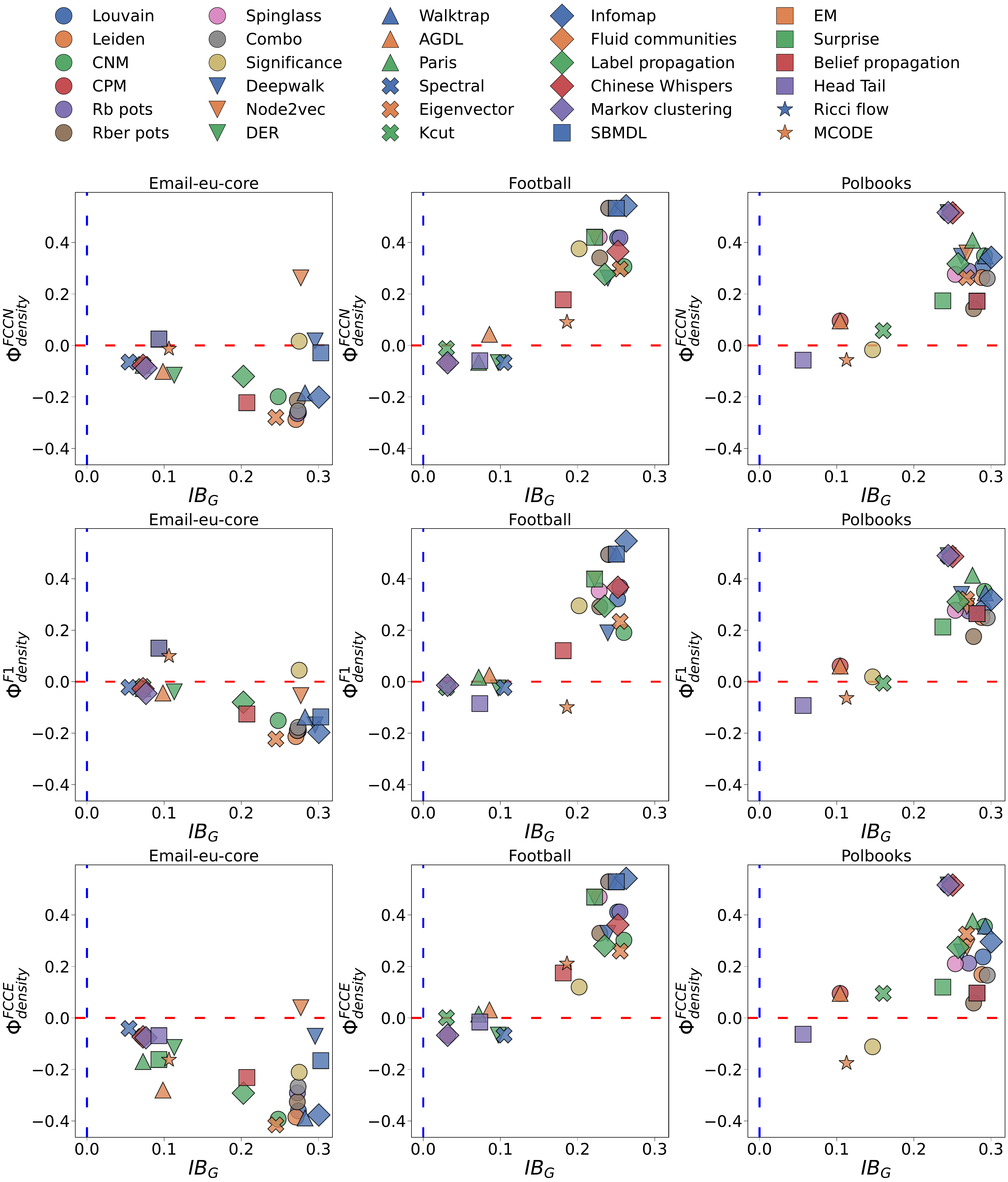}
    \caption{$IB_G$ measure against $\Phi_{density}$ measures for real-world networks. The red dotted line is the perfect fairness for $\Phi$, the blue dotted line is the perfect fairness for $IB_G$.}
    \label{fig:RW_IVP_density}
\end{figure}

Figure~\ref{fig:RW_IBg_vs_performance} presents the results of $IB_G$ against the quality measures from~\ref{subsec:cd_scores} on real-world networks mentioned in Table~\ref{tab:real_world_data}. We then analyse the $IB_G$ vs. the group-fairness measure $\Phi$, as shown in Figures~\ref{fig:RW_IVP_size},~\ref{fig:RW_IVP_conductance},~\ref{fig:RW_IVP_density}.

On real-world networks, for modularity, there is a clear trade-off: algorithms with the highest modularity tend to be the most individually unfair. This difference is expected, as real-world ``ground truth’’ labels are typically affected by external factors rather than well-separated structural communities, reducing the kind of explicit structure seen in ABCD graphs at low~$\xi$. Modularity scores are consistently highest on the football network, whose community structure is well defined and whose communities have similar sizes.
NMI shows similar behaviour to the Modularity, where the NMI is higher for football and polbooks, achieving higher clustering quality than email-Eu-core. Several algorithms reach NMI values around~0.92 on football (e.g., Significance, Surprise, Spinglass, Node2vec). Significance is particularly noteworthy, combining high NMI ($0.92$) with moderate fairness ($IB_G = 0.20$). Head Tail behaves differently from its behaviour on synthetic data, achieving competitive NMI scores while consistently maintaining $IB_G < 0.10$.

ARI shows different results than both modularity and NMI, and algorithms that appear strong under modularity or NMI, such as Paris, AGDL, Kcut, and Spectral, often dropped performance under ARI. NF1 results largely mirror ARI, especially on polbooks. On email-Eu-core, almost all algorithms achieve very low NF1 values, with Head Tail as the exception ($NF1 = 1.85$).

Figures~\ref{fig:RW_IVP_size},~\ref{fig:RW_IVP_conductance},~\ref{fig:RW_IVP_density} plot $IB_G$ against the group-fairness measures $\Phi_\text{size}$, $\Phi_\text{conductance}$, and $\Phi_\text{density}$ (with FCCN, F1, and FCCE variants). The red dotted line indicates perfect group fairness (under $\Phi$), and the blue dotted line indicates perfect individual fairness. Across real-world networks, most algorithms display systematic group-level biases: they generally favour larger communities, lower-conductance communities, and either sparser (email-Eu-core) or denser (football, polbooks) communities, depending on the network. A few methods, such as Head Tail, MCODE, Significance, and Markov Clustering, show consistent group fairness in specific settings, with Markov Clustering and Kcut being among the few that achieve fairness on both individual and group levels in the football network. However, even when algorithms behave group-fairly (e.g., Head Tail, Paris, Spectral, Chinese Whispers), they often fail to achieve low $IB_G$, especially on polbooks where no method shows meaningful individual fairness.

\section{Conclusion}\label{sec:conclusion}

%In this work, we introduced a novel fairness measure for Community Detection in the form of two complementary measures, $IB$ and $IB_G$, that quantify individual unfairness at the node and graph levels, respectively. The method is based on a Community Co-occurrence matrix capable of capturing the quantitative effects of contextual variation. We evaluated these measures across a diverse set of CD algorithms on both synthetic graphs and real-world networks, demonstrating their ability to reliably capture context variation and quantify disparities in community assignment outcomes. Experiments on ABCD networks demonstrated that individual unfairness has no correlation with group fairness or clustering accuracy, highlighting the distinction between the individual and group points of view on fairness. We further observed that no algorithm or specific community identification approach (e.g., modularity-based, probabilistic, spectral) is consistently fair across all datasets or fairness dimensions. Finally, the distributional analysis of $IB$ highlighted the mechanisms behind fairness failures, whether arising from over-segmentation, uniform misclassification, or uneven treatment of subsets of nodes.

%With growing attention to fairness in network science, individual-level fairness remains largely unexplored. 
In this work, we introduced novel individual fairness measures for community detection, addressing a significant gap in the literature. We developed two complementary measures, $IB$ and $IB_G$, that quantify individual unfairness at the node and graph levels, respectively, using a co-occurrence matrix, i.e., capable of capturing the numerical effects of contextual variation. We evaluated these measures across a diverse set of CD algorithms on both synthetic graphs and real-world networks, demonstrating their ability to reliably capture context variation and quantify disparities in community assignment outcomes. Experiments on ABCD networks show that CD methods can exhibit substantial individual unfairness despite achieving high group fairness, underscoring the importance of evaluating and designing methods with both individual and group fairness in mind. We further observed that no algorithm or specific community identification approach (e.g., modularity-based, probabilistic, spectral) is consistently fair across all datasets or fairness dimensions. Finally, the distributional analysis of $IB$ highlighted the mechanisms behind fairness failures, whether arising from over-segmentation, uniform misclassification, or uneven treatment of subsets of nodes.
%Experiments on ABCD networks demonstrated that CD methods can exhibit high individual unfairness even if they have high group fairness, highlighting to consider both individual and group fairness while designing and testing CD methods. 

%Experiments on ABCD networks demonstrated that individual unfairness can occur even when group fairness or clustering accuracy is high, underscoring that individual and group fairness are not interchangeable. 

In future, one can extend the proposed measures by incorporating node similarity, enabling their application to heterogeneous graphs where individuals differ in attributes. Moreover, a natural next step is to integrate the proposed individual fairness metrics, $IB$ and $IB_G$, into fairness-aware CD algorithms, thereby advancing the design of methods that are intrinsically guided by principles of individual fairness.

\bibliography{sn-bibliography}

\end{document}